\documentclass[12pt]{article}
\usepackage[margin=1.1in]{geometry}
\usepackage{float}

\usepackage{enumerate}
\usepackage{amsmath}
\usepackage{amssymb}
\usepackage{amsthm}
\usepackage{color}
\usepackage{xcolor}
\usepackage{graphicx}
\usepackage{bbm}
\usepackage{color}
\usepackage{geometry}
\usepackage{bbding}
\usepackage{ dsfont }
\usepackage{caption}
\usepackage{stackrel}
\usepackage{graphicx,xcolor}
\usepackage{tikz}
\usepackage{hyperref}
\usepackage{enumerate}
\usepackage{multicol}
\usepackage{soul}
\usepackage{tabularx}
\usepackage{multirow}
\usepackage{mdwlist}
\usepackage{enumitem}
\usepackage{accents}
\usepackage{mathtools}
\usepackage{subcaption}
\usepackage{natbib}
\newcommand*{\QEDA}{\hfill\ensuremath{\square}}
\makeatletter
\DeclareRobustCommand\widecheck[1]{{\mathpalette\@widecheck{#1}}}
\def\@widecheck#1#2{%
    \setbox\z@\hbox{\m@th$#1#2$}%
    \setbox\tw@\hbox{\m@th$#1%
       \widehat{%
          \vrule\@width\z@\@height\ht\z@
          \vrule\@height\z@\@width\wd\z@}$}%
    \dp\tw@-\ht\z@
    \@tempdima\ht\z@ \advance\@tempdima2\ht\tw@ \divide\@tempdima\thr@@
    \setbox\tw@\hbox{%
       \raise\@tempdima\hbox{\scalebox{1}[-1]{\lower\@tempdima\box
\tw@}}}%
    {\ooalign{\box\tw@ \cr \box\z@}}}
\makeatother

\usepackage{threeparttable}
\newcommand{\deleq}{\stackrel{\Delta}{=}}
\newtheorem{example}{Example}
\newtheorem{proposition}{Proposition}
\newtheorem{theorem}{Theorem}
\newtheorem{lemma}{Lemma}

\DeclareMathOperator*{\argmin}{arg\,min}
\usepackage{mathtools}

\makeatletter
\newenvironment{taggedsubequations}[1]
 {%
  \addtocounter{equation}{-1}%
  \begin{subequations}%
  \def\@currentlabel{#1}%
  \renewcommand{\theequation}{#1.\arabic{equation}}%
 }
 {\end{subequations}}
\makeatother 
\newcommand{\RN}[1]{
  \textup{\uppercase\expandafter{\romannumeral#1}}}
\newcommand{\Lambdag}{\Lamb^{\dagger}}

\newcommand{\Tj}{\T^{\j}}

\newcommand{\K}{\mathcal{K}}
\renewcommand{\(}{\left(}
\renewcommand{\)}{\right)}
\newcommand{\tI}{\t^{I}}

\newcommand{\Lamb}{\Lambda}
\newcommand{\titil}{\tilde{t}^{\i}}
\newcommand{\tmitil}{\tilde{\t}^{-\i}}
\newcommand{\tmihat}{\widehat{\t}^{-\i}}
\newcommand{\tihat}{\widehat{\t}^{\i}}
\newcommand{\that}{\widehat{\t}}
\newcommand{\tj}{\t^j}
\newcommand{\qtjr}{\q^{j}_\r (\tj)}
\newcommand{\qtirtil}{\q^\i_\r(\titil)}
\newcommand{\tjtil}{\tilde{\t}^j}
\newcommand{\qtjrtil}{\q^j_\r(\tjtil)}
\renewcommand{\j}{j}

\newcommand{\lambj}{\lamb^{\j}}
\newcommand{\ldagij}{\l^{\i \j, \dagger}}
\newcommand{\ldagijtil}{\tilde{\l}^{\i \j, \dagger}}
\newcommand{\Ldagij}{\F^{\i \j, \dagger}}
\newcommand{\Lambmimj}{\Lamb^{-\i \j}}
\newcommand{\wdagij}{\w^{\i \j, \dagger}}

\newcommand{\zij}{z^{\i\j}}
\newcommand{\alti}{\mu^{\ti}}
\newcommand{\betir}{\nu_{r}^{\ti}}

\newcommand{\lag}{\mathcal{L}}
\newcommand{\pro}{\mathrm{Pr}}

\newcommand{\ps}{\theta}

\newcommand{\md}{J}
\newcommand{\mdi}{J^i}

\newcommand{\game}[0]{\Gamma(\lamb)}

\newcommand{\playerSet}[0]{\mathcal{I}}
\newcommand{\I}{\playerSet}
\newcommand{\player}[0]{i}
\renewcommand{\i}{\player}

\newcommand{\playerTypeSet}[0]{\mathcal{T}} 
\newcommand{\playerType}[0]{t} 
\renewcommand{\t}{\playerType}
\newcommand{\tone}{\t^1}
\newcommand{\ttwo}{\t^2}

\newcommand{\T}{\playerTypeSet}
\newcommand{\Tone}{\playerTypeSet^1}
\newcommand{\Ttwo}{\playerTypeSet^2}

\newcommand{\ti}{t^i}
\newcommand{\Tmi}{\playerTypeSet^{-\player}}
\newcommand{\tmi}{\playerType^{-\player}}

\newcommand{\belief}{\beta}
\newcommand{\muti}{\belief^i(\s, \tmi|\ti)}
\newcommand{\mutj}{\belief^j(\s, \t^{-j}|\t^j)}

\newcommand{\lamb}{\lambda}
\newcommand{\lambi}{\lamb^{\i}}

\newcommand{\Reone}{\Lambda_1}
\newcommand{\Retwo}{\Lambda_2}
\newcommand{\Rethree}{\Lambda_3}
\newcommand{\Reoneij}{\Lambda_1^{\i\j}}
\newcommand{\Retwoij}{\Lambda_2^{\i\j}}
\newcommand{\Rethreeij}{\Lambda_3^{\i\j}}

\newcommand{\ecrtjwe}{\mathbb{E}[c_\r(\qwe)|\tj]}

\newcommand{\ldag}{\l^{\dagger}}

\newcommand{\Ldag}{\F^{\dagger}}
\newcommand{\lambl}{\underline{\lamb}^1}
\newcommand{\lambli}{\underline{\lamb}^\i}

\newcommand{\lamblip}{\underline{\underline{\lamb}}^{\i}}

\newcommand{\lambup}{\bar{\lamb}^1}
\newcommand{\lambupi}{\bar{\lamb}^\i}

\newcommand{\Dplambij}{\nabla_{z^{\i \j}} \Plamb}

\newcommand{\lambupip}{\bar{\bar{\lamb}}^{\i}}

\newcommand{\totalDemand}{\mathrm{D}}
\newcommand{\D}{\totalDemand}

\newcommand{\load}{f}
\renewcommand{\l}{\load}
\newcommand{\lwe}{\l^{*}}
\newcommand{\lwer}{\l^{*}_{\r}}
\newcommand{\lr}{\load_{\path}}
\newcommand{\lrwe}{\l_r^{*}}
\newcommand{\we}{w_{\e}}
\newcommand{\w}{w}

\newcommand{\wwelamb}{w^{*}(\lambda)}
\newcommand{\wdag}{w^{\dagger}}
\newcommand{\wwe}{w^{*}}

\newcommand{\stateSet}[0]{\mathcal{S}}
\newcommand{\state}[0]{s}
\newcommand{\s}{\state}
\renewcommand{\S}{\stateSet}
\renewcommand{\a}{\mathbf{a}}
\newcommand{\n}{\mathbf{n}}

\newcommand{\potential}{\Phi}

\newcommand{\E}[0]{\mathcal{E}}
\newcommand{\e}[0]{e}

\newcommand{\routeSet}[0]{\mathcal{R}}
\newcommand{\R}{\routeSet}
\renewcommand{\path}[0]{r}
\renewcommand{\r}{\path}

\newcommand{\F}{\mathcal{F}}

\newcommand{\stratSet}[0]{\mathcal{Q}}
\newcommand{\strat}[0]{q}
\newcommand{\q}{\strat}
\newcommand{\Q}{\stratSet}
\newcommand{\Qi}{\stratSet^{\i}}
\newcommand{\qi}{\strat^{\player}}

\newcommand{\qtir}{\strat_{\path}^{\player}(\ti)}

\newcommand{\qone}{\strat^1}
\newcommand{\qtwo}{\strat^2}

\newcommand{\qweone}{\strat^{1*}}
\newcommand{\qwetwo}{\strat^{2*}}

\newcommand{\qwe}{\q^{*}}
\newcommand{\qwetir}{\q^{\i*}_{\r}(\ti)}

\newcommand{\suplamb}{\bar{\I}(\lamb^{\dagger})}

\newcommand{\cfun}{c}
\newcommand{\cfunset}{\mathcal{C}}
\newcommand{\cesfun}{\cfun_{\e}^{\s}}

\newcommand{\ecrti}{\mathbb{E}[\cfun_{\r}({\q})|\ti]}
\newcommand{\ecrtiwe}{\mathbb{E}[\cfun_{\r}({\qwe})|\ti]}
\newcommand{\ecrpti}{\mathbb{E}[\cfun_{\r^{'}}({\q})|\ti]} 
\newcommand{\Ti}{\T^{\i}}

\newcommand{\ecrptiwe}{\mathbb{E}[\cfun_{\r'}(\q^{*})|\ti]}

\newcommand{\cilamb}{C^{\i*}(\lamb)}
\newcommand{\cjlamb}{C^{\j*}(\lamb)}
\newcommand{\rvlambij}{V^{\i\j*}(\lamb)}

\newcommand{\conelamb}{C^{1*}(\lamb)}
\newcommand{\ctwolamb}{C^{2*}(\lamb)}

\newcommand{\aonea}{\alpha_1^{\a}}
\newcommand{\aonen}{\alpha_1^{\n}}
\newcommand{\atwo}{\alpha_2}

\newcommand{\Plamb}{\Psi(\lamb)}

\newcommand{\Lwe}{\mathcal{F}^{*}}

\newcommand{\Qwe}{\mathcal{\Q}^{*}}

\newcommand{\tonehat}{\widehat{\t}^1}
\newcommand{\ttwohat}{\widehat{\t}^2}

\newcommand{\tmjhat}{\widehat{\t}^{-\j}}

\newcommand{\mdl}{\widehat{\md}}

\newcommand{\Lwelamb}{\F^{*}(\lamb)}

\newcommand{\wpotential}{\widecheck{\potential}}
\newcommand{\lpotential}{\widehat{\potential}}

\newcommand{\ctwo}{c_2}
\newcommand{\lone}{\l_{1}}
\newcommand{\ltwo}{\l_{2}}
\newcommand{\rone}{\r_1}
\newcommand{\rtwo}{\r_2}

\newcommand{\lambmimj}{\lamb^{-\i \j}}
\renewcommand{\k}{k}
\newcommand{\lambk}{\lamb^{\k}}

\newcommand{\abar}{\bar{\alpha}_1}

\begin{document}

\title{Value of Information in Bayesian Routing Games}
\author{Manxi Wu, Saurabh Amin, and Asuman E. Ozdaglar
\thanks{M. Wu is with the Institute for Data, Systems, and Society, S. Amin is with the Laborotary for Information and Decision Systems, and A. Ozdaglar is with the Department of Electrical Engineering and Computer Science, Massachusetts Institute of Technology (MIT), Cambridge, MA, USA
        {\tt\small \{manxiwu,amins,asuman\}@mit.edu}}%
}
\date{}
\maketitle
\begin{abstract}
    We study a routing game in an environment with multiple heterogeneous information systems and an uncertain state that affects edge costs of a congested network. Each information system sends a noisy signal about the state to its subscribed traveler population. Travelers make route choices based on their private beliefs about the state and other populations' signals. The question then arises, ``How does the presence of asymmetric and incomplete information affect the travelers' equilibrium route choices and costs?'' We develop a systematic approach to characterize the equilibrium structure, and determine the effect of population sizes on the relative value of information (i.e. difference in expected traveler costs) between any two populations. This effect can be evaluated using a population-specific size threshold. One population enjoys a strictly positive value of information in comparison to the other if and only if its size is below the corresponding threshold. We also consider the situation when travelers may choose an information system based on its value, and characterize the set of equilibrium adoption rates delineating the sizes of subscribed traveler populations. The resulting routing strategies are such that all travelers face an identical expected cost and no traveler has the incentive to change her subscription.
    
    \vspace{0.2cm}
    \noindent\textbf{Index Term}: Games/group decisions: Nonatomic; Transportation: Technology; Information systems.
\end{abstract}

\section{Introduction}\label{sec:introduction}
Travelers are increasingly relying on traffic navigation services to make their route choice decisions. In the past decade, numerous services have come to the forefront, including Waze/Google maps, Apple maps, INRIX, etc. These Traffic Information Systems (TISs) provide their subscribers with costless information about the uncertain network condition (state), which is typically influenced by exogenous factors such as weather, incidents, and road conditions. The information provided by TIS can be especially useful in making travel decisions when a change in state corresponds to changes in travel times of multiple edges or routes of the network. Experiential evidence suggests that the accuracy levels of TISs are less than perfect, and exhibit heterogeneities due to the inherent technological differences in data collection and analysis procedures. Moreover, travelers may use different TISs or choose not to use them at all, depending on factors such as marketing, usability, and availability. Therefore, we can reasonably expect that travelers face an environment of asymmetric and incomplete information about the network state. 

Importantly, information heterogeneity can directly influence the travelers' route choice decisions, and the resulting congestion externalities. Consider an example where some travelers are informed by their TIS that a certain route has an incident. Taking a detour based on this information may not only reduce their own travel time, but also benefit the uninformed travelers by shifting traffic away from the affected route. However, if too many travelers take the detour, then this alternate route will also start getting congested, limiting the benefits of information (Section \ref{motivate}). Thus, the question arises as to how information heterogeneity impacts the travelers’ route choices and costs. 

In this article, we develop a game-theoretic approach to study this question. We consider a routing game in which the travelers are privately informed about the network state by their respective TIS, and choose strategies based on their beliefs about the state and other travelers' behavior. We conduct a systematic study of how the sizes of heterogeneously informed traveler populations affect the equilibrium structure and the relative value of information faced by travelers subscribed to one TIS in comparison to another TIS. Furthermore, we characterize the set of equilibrium adoption rates when travelers can choose any available TIS based on its relative value.

\subsection{Our Model and Contributions}
We model the traffic routing problem in an asymmetric and incomplete information environment as a Bayesian routing game. We consider a general traffic network with an uncertain state that is realized from a finite set according to a prior probability distribution. The cost function (travel time) of each edge in the network is increasing in the aggregate traffic load on that edge. Moreover, the edge costs are \emph{state-dependent} in that the state can affect them in various ways. 
There are multiple heterogeneous TISs, each sending a noisy signal of the state to its subscribed traveler population. The signal sent by each TIS is privately known to only its subscribers, but the joint distribution of the state and all signals is known by all travelers (common knowledge). This joint distribution is the \emph{common prior} of the game, and each population's private belief of the state and other populations’ private signals is derived from it. Our information environment is general in that we do not impose any structural assumption on the common prior. For example, the signals of different TISs can be correlated or independent, conditional on the state. Also, one TIS can be more accurate than another in some states, but less in the other states. Thus, we do not assume that the TISs are ordered according to the accuracy of their signals (Section \ref{model_environment}). 

We use \textsl{Bayesian Wardrop Equilibrium (BWE)} as the solution concept of our game. In a BWE, all populations assign demand on routes with the smallest expected cost based on their private beliefs. In fact, our game is a weighted potential game, i.e. the set of BWE is the optimal solution set of a convex optimization problem that minimizes the weighted potential function over the set of feasible routing strategies of traveler populations (Theorem \ref{q_opt}). This property establishes the essential uniqueness of BWE, i.e. the equilibrium edge load vector is unique. However, the strategy-based optimization problem is not directly useful for analyzing how the set of equilibrium strategies and population costs depend on the population size vector. To address this issue, we provide a characterization of the set of feasible route flows (i.e. route flows induced by feasible strategy profiles) as a convex polytope (Proposition \ref{Lprime}), and show that this set is the optimal solution set of another convex optimization problem (Proposition \ref{l_opt}). The optimal value of this flow-based formulation is the value of weighted potential function in equilibrium (Section \ref{Sec:potential_game}). 

The flow-based formulation enables us to analyze the sensitivity of equilibrium structure with respect to perturbations in the size vector. The constraints in this formulation include: basic route feasibility constraints, a set of size-independent equality constraints, and a set of size-dependent inequality constraints. The equality constraints ensure that any shift in route flows resulting from a change of signal received by one population is independent of the signals received by other populations. Each inequality constraint corresponds to a single population, and ensures that the maximum extent to which the received signals impact its equilibrium routing behavior is limited by the population's size. Hence, we refer to these inequality constraints as \emph{information impact constraints} (IICs). Consequently, the effects of perturbation in the size vector on the equilibrium structure can be studied by evaluating the tightness of the IICs corresponding to the perturbed populations at the optimum of the flow-based formulation. 

In particular, Theorem \ref{l_behavior} describes how the qualitative properties of equilibrium route flows change under perturbations in the sizes of any two populations, with sizes of all other populations being fixed (i.e. directional perturbations of the size vector). Among the two perturbed populations, we say that one population is the ``minor population'' if its size is smaller than a certain (population-specific) threshold. Then, the corresponding IIC is tight in equilibrium, i.e. the impact of information on the minor population is fully attained. These population-specific thresholds depend on the common prior distribution as well as sizes of all other populations. In fact, each threshold can be computed by solving a linear program. Based on the two thresholds, we can distinguish three qualitatively distinct equilibrium regimes: In the two side regimes exactly one population assumes the minority role, on the other hand, in the middle regime neither population is minor. 

We can apply Theorem \ref{l_behavior} to analyze the sensitivity of the equilibrium value of the weighted potential function under directional perturbation of population sizes (Proposition \ref{bathtub}). In the middle regime, perturbing the relative sizes does not change the equilibrium value of the potential function. On the other hand, in the two side regimes the value of the potential function monotonically decreases as the size of the minor population increases. Thanks to the essential uniqueness of BWE, the equilibrium edge load vector does not change with this directional perturbation if and only if the size vector falls in the middle regime. 

These results allow us to compare the expected cost in equilibrium faced by travelers in any pair of populations. In particular, we can evaluate how this cost difference --- which we call the \emph{relative value of information} --- changes with pairwise size perturbations. By using the results on sensitivity analysis of general convex optimization problems (\citet{fiacco1983introduction}, \citet{rockafellar1984directional}), we show that the relative value of information is proportional to the derivative of the equilibrium value of potential function in the direction of perturbation. Combining this observation with Proposition \ref{bathtub}, we obtain that the minor population faces a lower cost relative to the other population in the two side regimes; whereas both populations face identical costs in the middle regime. Importantly, the relative value of information is non-increasing in the size of its subscribed population (Theorem \ref{prop:relative_value}). This result is based on the intuition that an individual traveler faces higher congestion externality when more travelers have access to the same information, and hence make their route choices according to the same strategy. Thus, an increase in the size of minor population decreases the relative imbalance in congestion externality, thereby reducing the advantage enjoyed by its travelers over the other population.  

Our results can be easily specialized to a simpler information environment in which one of the populations does not have an access to TIS (uninformed population). In this case, we obtain that the equilibrium cost of the uninformed population is no less than that of any other population regardless of the size vector (Proposition \ref{no_info}). That is, having access to a TIS always leads to a non-negative relative value of information in comparison to being uninformed (Section \ref{equilibrium_regime}). 

Furthermore, we extend our approach of comparing the relative value of information for any pair of populations to study how the equilibrium outcome depends on the population sizes in general. In particular, we characterize a non-empty set of size vectors, for which the equilibrium edge load vector is size-independent, and the value of potential function is minimized (Proposition \ref{theorem:intermediate}). This result enables us to study situations when travelers can choose their TIS subscription based on its value. Then, the notion of adoption rate of TIS becomes relevant, because the value of information to users (as well as non-users) of a TIS depends on the size of the population with access to information provided by it. Intuitively, if it is possible for the non-users of a particular TIS to receive a greater value by adopting it, then they may continue to do so until there is no longer a positive relative value in adopting that TIS. Theorem \ref{order_population} shows that the set of equilibrium adoption rates (i.e. fraction of travelers choosing each TIS in equilibrium) is exactly the set of population sizes that minimize the value of potential function at equilibrium. Naturally, any size vector that is induced by travelers' equilibrium choice of TIS satisfies the property that all travelers face identical costs and no traveler has an incentive to change her TIS subscription (Section \ref{sec:general_cost}).


We conclude our work in Section \ref{concluding}. We include supplementary results, examples, and most of the proofs in the electronic companion of this article.



\subsection{Related Work}\label{related_work}
\emph{Congestion games}. Well-known results in classical congestion games include their equivalence with potential games (\citet{rosenthal1973class}, \citet{monderer1996potential}, and \citet{sandholm2001potential}), analysis of network formation games as congestion games (\cite{gopalakrishnan2014potential}, and \cite{tardos2007network}), and equilibrium inefficiency (\citet{roughgarden2004bounding}, \citet{Koutsoupias}, \cite{correa2007fast}, \cite{acemoglu2007competition}, and \cite{nikolova2014mean}). \cite{Milchtaich1996111} has studied congestion games with player-specific cost functions, which can model both heterogeneous private information and heterogeneous preferences. However, the existence of potential function is not guaranteed with general player-specific cost functions. In our game, the heterogeneity in the expected costs among populations arises only due to heterogeneous private beliefs, which are derived from a common prior. This feature makes our game a weighted potential game (\cite{sandholm2001potential}). Moreover, \cite{acemoglu2018informational} studied a congestion game where travelers have different information sets about the available edges (routes). While their work focuses on heterogeneous information about the network structure, we use a Bayesian approach to model the information heterogeneity resulting from the differences in TIS access and accuracy.

\emph{Traffic Information Systems}. Prior work has studied the effects of TIS on travelers' departure time choices (\cite{arnott1991does}, and \cite{khan2018bottleneck}), and on their route choices (\cite{ben1991dynamic}, \cite{ben1996impact}). In particular, the paper \cite{mahmassani1991system} used simulation approach to show that as more travelers receive information, the informed travelers gradually start facing higher costs and their relative value of being informed diminishes. Our analysis, when applied to the game with one uninformed population and other more informed ones, also leads to similar conclusion. Our results are more general because they are applicable to routing games with multiple heterogeneously informed populations with arbitrary TIS accuracies.   


\emph{Value of Information}. In a classical paper, \cite{blackwell1953equivalent} showed that for a single decision maker, more informative signal always results in higher expected utility. 
In game-theoretic settings, it is generally difficult to determine whether the value of information in equilibrium for individual players and/or society is positive, zero or negative (see \cite{hirshleifer1971private}, and \cite{haenfler2002social}). However, the value of information is guaranteed to be positive when certain conditions are satisfied; see for example \cite{neyman1991positive}, and \cite{lehrer2006restrictions}. Since travelers are non-atomic players in our game, the relative value of information between any two TISs is equivalent to the value of information for an individual traveler when her subscription changes unilaterally. We give precise conditions on the population sizes under which the value of information in our Bayesian routing game is positive, zero, or negative.

\section{Motivating Example}\label{motivate}
In this section, we motivate our analysis using a simple game of two asymmetrically informed traveler populations routing over a network of two parallel routes, denoted $\rone$ and $\rtwo$. The network state $\s$ belongs to the set $\S=\{\a, \n\}$, where the state $\a$ represents an incident condition on $\rone$, and the state $\n$ represents the nominal condition. The state $\a$ occurs with probability $p \in (0, 1)$. The network faces a unit size of demand ($\D=1$), which is comprised of two traveler populations: population 1 with size $\lamb^1$ and population 2 with size $\lamb^2=1-\lamb^1$. Each population $\i \in \I=\{1,2\}$ receives a signal $\ti \in \Ti=\{\a, \n\}$ from its TIS. 
Assume for simplicity that population 1 receives the correct state with probability 1 (i.e. complete information), and population 2 receives signal $\a$ or $\n$ with probability $0.5$, independent of the state (i.e. no information). 

Let $\qi_1(\ti)$ denote the traffic demand assigned to route $\rone$ by population $\i$ when receiving signal $\ti$; the remaining demand $\(\lamb^i-\qi_1(\ti)\)$ is assigned to route $\rtwo$. Since the signal $\ttwo$ is independent of the state, we have $\qtwo_1(\a)=\qtwo_1(\n) \deleq \qtwo_1$. A feasible demand assignment must satisfy the constraints: $0 \leq \qone_1(\tone) \leq \lamb^1 $ and $0 \leq \qtwo_1 \leq \lamb^2$. For this example, we can represent a routing strategy profile as $q_1=\(\qone_1(\a), \qone_1(\n), \qtwo_1\)$. We denote the aggregate route flow on $\r$ as $\lr$. We consider simple affine cost functions: the cost of $r_1$ is $c_1^{\a}(\lone)=\aonea \lone +b$ in state $\a$, and $c_1^{\n}(\lone)=\aonen \lone +b$ in state $\n$; the cost of $r_2$ is $\ctwo(\ltwo)=\atwo \ltwo+b$. The slopes satisfy $\aonen< \atwo<\aonea$. Since population 1 has complete information, its travelers know the exact cost function in both states. However, since population 2 travelers are uninformed, they make their route choices based on the expected cost of each route, evaluated according to the prior distribution of states. 

This routing game with heterogeneously informed traveler populations admits a Bayesian Wardrop equilibrium, as discussed in Section \ref{model_environment}. Let $\qwe_1=\(\qweone_1(\a), \qweone_1(\n), \qwetwo_1 \)$ denote an equilibrium strategy profile. Each population, given the signal it receives, can either assign all its demand on one of the two routes, or splits on both routes. Thus, there are $3^3=27$ possible cases. Our results can be used to study how the equilibrium strategies and route flows change as the size of a population varies from 0 to 1. Detailed analysis for this simple routing game and some interesting variants are available in \cite{wu2017informational}. Specifically, we find that there exists a threshold size of population 1, $0 < \lambl=\atwo\(\frac{1}{\aonen+\atwo} -\frac{1}{\aonea+\atwo}\) < 1$, such that the qualitative structure of equilibrium routing strategies is different based on whether $\lamb^1 \in [0, \lambl)$ or $\lamb^1 \in [\lambl, 1]$. 

In the first regime, i.e. when $\lamb^1 \in [0, \lambl)$, the game admits a unique equilibrium: $\qweone_1(\a)=0$, $\qweone_1(\mathbf{n})=\lamb^1, \text{ and } \qwetwo_1=\frac{\atwo}{\abar+\atwo}-\lamb^1\frac{(1-p) (\aonen+\atwo)}{\abar+\atwo}$, where $\abar=p\aonea+(1-p)\aonen$. This equilibrium regime corresponds to the following outcome: in state $\n$ (resp. state $\a$), population 1 assigns all its demand on route $\rone$ (resp. route $\rtwo$), and population 2 splits its demand on both routes. The induced equilibrium flow on $\rone$ is given by $\l_1^{*}=\qwetwo_1$ if $\tone=\a$, and $\l_1^{*}=\lamb^1+\qwetwo_1$ if $\tone=\n$. 

On the other hand, in the second regime, i.e. when $\lamb^1 \in [\lambl, 1]$, the equilibrium set may not be singleton, and can be represented as follows: $\qweone_1(\a)=\chi$, $\qweone_1(\mathbf{n})=\lambl+\chi, \text{ and }  \qwetwo_1=\frac{\atwo}{\aonea+\atwo}-\chi$, where $\max \left\{0, \lamb^1 -\frac{\aonea}{\aonea +\atwo}\right\} \leq \chi \leq \min \left\{\frac{\atwo}{\aonea+\atwo}, \lamb^1  -\lambl  \right\}$. Thus, the equilibrium set is a one-dimensional interval for $\lamb^1 \in (\lambl, 1)$, and a singleton set for $\lamb^1=\lambl$ or $\lamb^1=1$. In this regime, both populations face identical expected route costs in equilibrium. Consequently, each population splits its demand on both routes. Moreover, the equilibrium route flow on each route is unique and independent of $\lamb^{1}$: $\l_1^*=\frac{\atwo}{\aonea+\atwo}$ if $\tone=\a$, and $
\l_1^*=\frac{\atwo}{\aonen+\atwo}$ if $\tone=\n$. 

Notice that when $\lamb^1 \in [0, \lambl)$, we have $\q_1^{1*}(\n)-\qweone_1(\a)=\lamb^1$, i.e. population 1 shifts all its demand to $\rtwo$ when receiving the signal about the incident on $\rone$. However, if $\lamb^1 \in [\lambl, 1]$, we have $\q_1^{1*}(\n)-\qweone_1(\a)=\lambl < \lamb^1$, i.e. the change in the received signal only influences a part of travelers in population 1. One can say that the information impacts the entire demand of population 1 in the first regime, but not in the second regime. 

For any feasible $\lamb =\(\lamb^{1}, \lamb^{2}\)$, we can calculate the equilibrium population costs, denoted $\cilamb$. If $\lamb^1 \in [0, \lambl)$, 
it is easy to check that $\ctwolamb-\conelamb>0$, i.e. when the state information is only available to a small fraction of travelers, the informed travelers have an advantage over the uninformed ones. On the other hand, if the size of informed population exceeds the threshold $\lambl$, then $C^{1*}(\lamb)=C^{2*}(\lamb)$. In this case, all travelers face identical cost in equilibrium. 
 
\begin{figure}[H]
    \begin{subfigure}[b]{0.32\textwidth}
        \includegraphics[width=\textwidth]{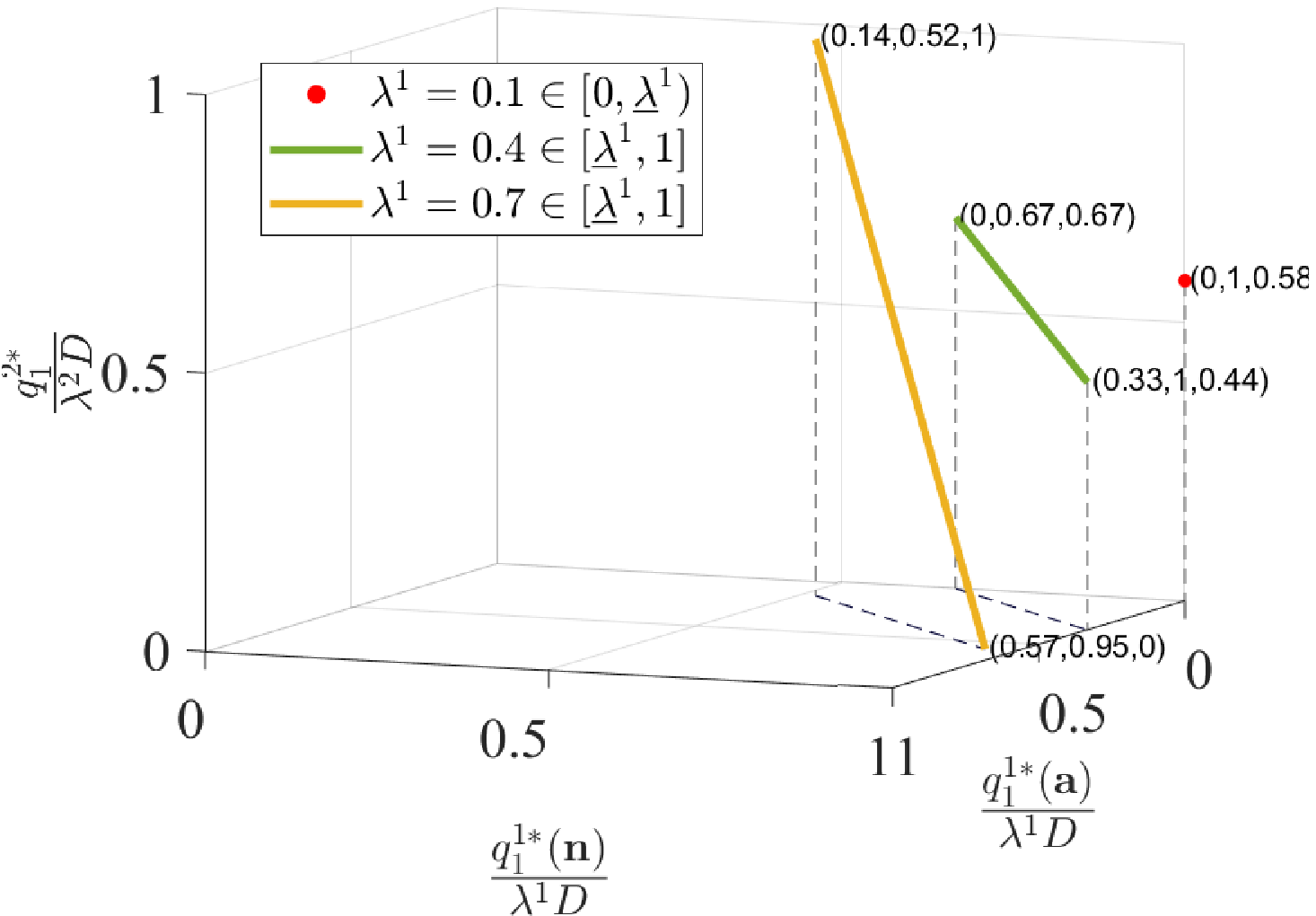}
        \caption{}
        \label{fig:H_100_L_50_D_1_q}
    \end{subfigure}
~
	\begin{subfigure}[b]{0.32\textwidth}
        \includegraphics[width=\textwidth]{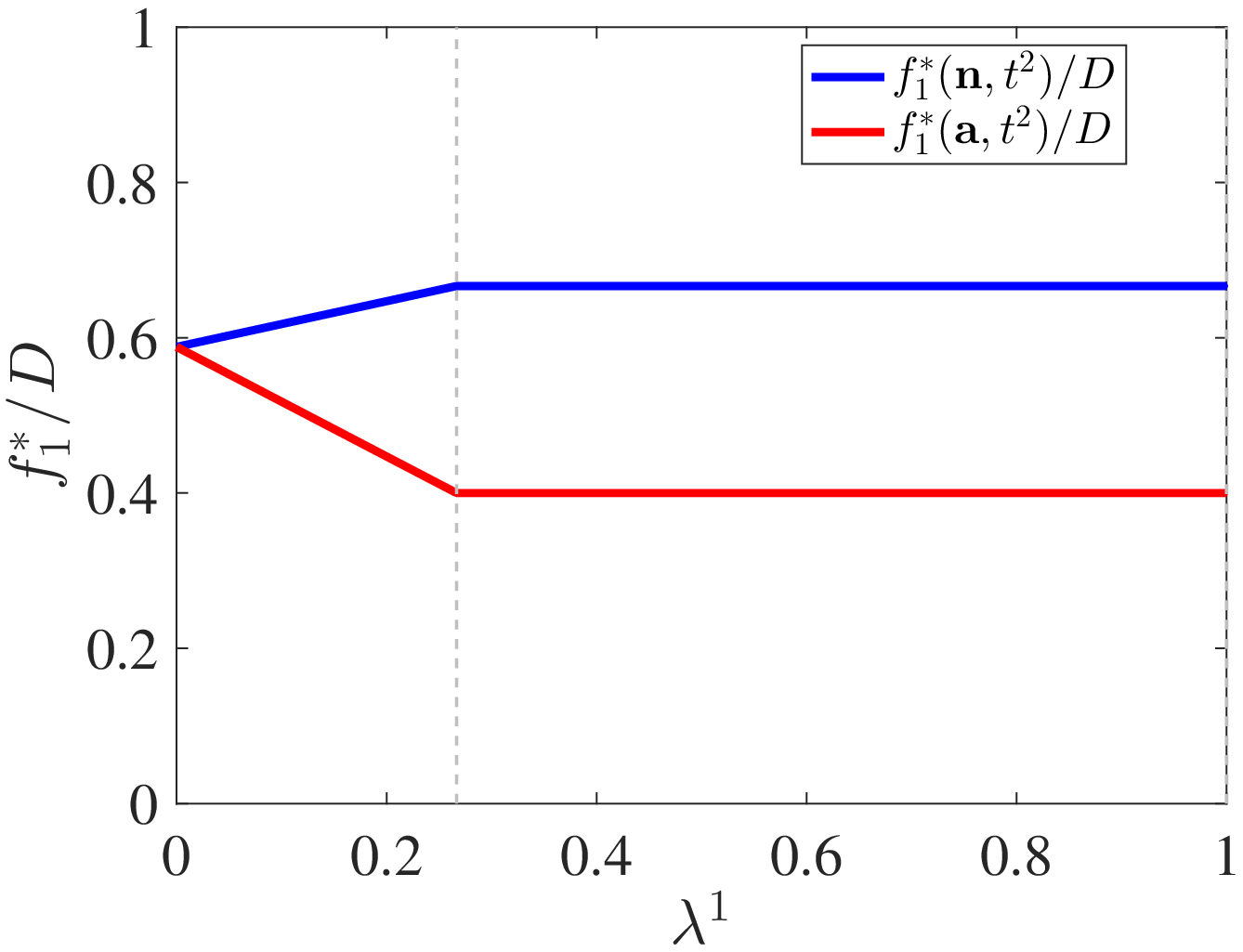}
        \caption{}
        \label{fig:H_100_L_50_D_1_f}
    \end{subfigure}
~
   \begin{subfigure}[b]{0.32\textwidth}
        \includegraphics[width=\textwidth]{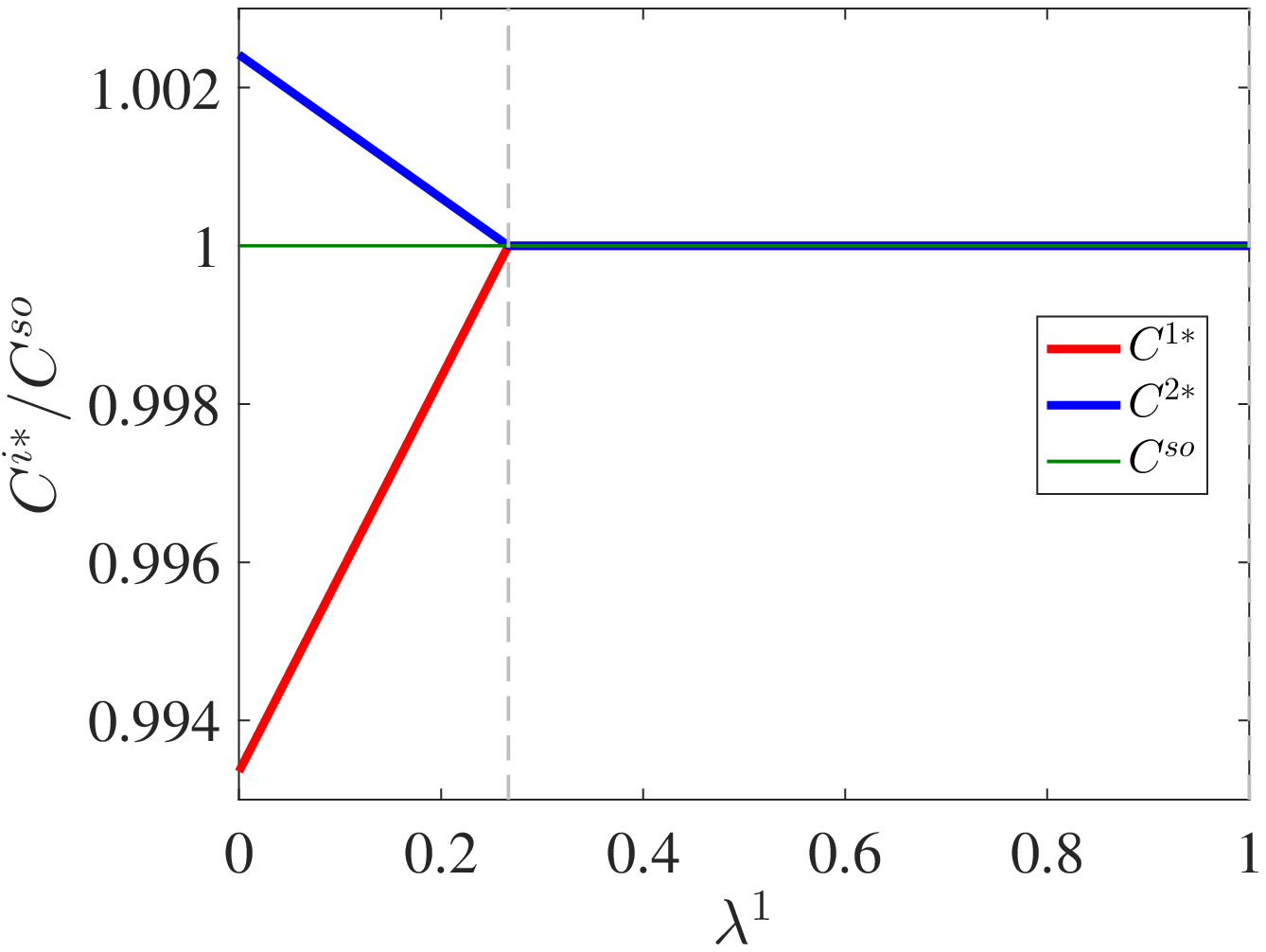}
        \caption{}
        \label{fig:H_100_L_50_D_1_p}
    \end{subfigure}

    \caption{Effects of varying population 1 size on equilibrium structure and costs: (a) Population strategies, (b) Flow on route $r_1$, (c) Population costs.}
    \label{fig:H_75_L_50_D_1_stra}
\end{figure}

We illustrate the aforementioned results in Fig. \ref{fig:H_75_L_50_D_1_stra} using the following parameters: $\aonen=1$, $\aonea=3$, $\atwo=2$, $b=20$, and $p=0.2$. The costs are normalized by the socially optimal cost, denoted $C^{so}$, which is the minimum cost achievable by a social planner with complete information of the state. Our subsequent analysis considers more general settings with multiple (heterogeneous and possibly correlated) information systems and general network topology.

\section{Model}\label{model_environment}                                                                                                                                                                                                                                                                                                                                                                    
\subsection{Environment} \label{network}
To generalize the simple routing game in Section \ref{motivate}, we consider a transportation network modeled as a directed graph. For ease of exposition, we assume that the network has a single origin-destination pair. All our results apply to networks with multiple origin-destination pairs; see Sec \ref{appendix_D}. Let $\E$ denote the set of edges and $\R$ denote the set of routes. The finite set of network states, denoted $\S$, represents the set of possible network conditions, such as incidents, weather, etc. The network state, denoted $\s$, is randomly drawn by a fictitious player ``Nature'' from $\S$ according to a distribution $\ps \in \Delta(\S)$, which determines the prior probability of each state. For any edge $\e \in \E$ and state $\s \in \S$, the state-dependent edge cost function $\cesfun(\cdot)$ is a positive, increasing, and differentiable function of the load through the edge $\e$. The state can impact the edge costs in various ways. 

The network serves a set of non-atomic travelers with a fixed total demand $\D$. We assume that each traveler is subscribed exclusively to one of the TIS in the set $\I=\{1, \cdots, I\}$. We refer to the set of travelers subscribed to the TIS $\i \in \I$ as population $\i$. All travelers within a population receive an identical signal from their TIS. Let $\lambi$ denote the ratio of population $\i$'s size and the total demand $\D$. We also consider degenerate situations when the sizes of one or more populations approach 0. Thus, a vector of population sizes $\lamb = \(\lamb^1, \dots, \lamb^{I}\)$ satisfies $\sum_{\i \in \I} \lambi =1$ and $\lambi \geq 0$ for any $\i \in \I$.
The size vector $\lamb$ is considered as given in our analysis of equilibrium structure and costs (Sections \ref{Sec:potential_game} and \ref{equilibrium_regime}). In Section \ref{sec:general_cost} we consider a more general situation where $\lamb$ results from the travelers' TIS subscription choices.

Each TIS $\i \in \I$ sends a noisy signal $\ti$ of the state to population $\i$. The signal received by each population determines its type (private information). We assume that the type space of population $\i$ is a finite set, denoted as $\Ti$. Note that the type spaces $\Ti$ and the state space $\S$ need not be of the same size. Let $\t \deleq \(\tone, \ttwo, \dots, \tI\)$ denote a type profile, i.e. vector of signals received by the traveler populations; thus, $\t \in \T\deleq \prod_{\i \in \I} \Ti$. The joint probability distribution of the state $\s$ and the vector of signals $\t$ is denoted $\pi \in \Delta(\S \times \T)$, and it is the common prior of the game. The marginal distribution of $\pi$ on states is consistent with the common prior, i.e. $\sum_{\t \in \T} \pi(\s, \t)=\ps(\s)$ for all $\s \in \S$. The conditional probability of type profiles $\t$ on the state $\s$ is given by $p(\t|\s)=\frac{\pi(\s, \t)}{\ps(\s)}$, i.e. the joint distribution of signals received by the populations when the network state is $\s$. In our modeling environment, the signals of different TIS can be correlated, conditional on the state. 

Each population $\i$ generates a belief about the state $\s$ and the other populations' types $\tmi$ based on the signal received from the information system $\i \in \I$. We denote the population $\i$'s belief as $\muti \in \Delta (\S \times \Tmi)$.

The routing strategy of each population $\i \in \I$ is a function of its type, denoted as $\qi(\ti) =\(\qtir\)_{\r \in \R}$. One way to describe the generation of routing strategies is that each TIS $\i\in\I$ sends a noisy 
signal $\ti$ of the state to its subscribed population, and the individual route choices of non-atomic travelers results in an aggregate routing strategy $\qi(\ti)$. An alternative viewpoint is that $\qi(\ti)$ is a direct result of strategy route recommendations sent by each TIS to its subscribed population. That is, each TIS $\i \in \I$ routes travelers in population $\i$ according to the function $\qi(\ti)$. For our purpose, these two viewpoints are equivalent in that given any population $\i \in \I$, and any type $\ti \in \Ti$, the demand of travelers on route $\r \in \R$ is $\qtir$.

We say that a routing strategy profile $\q \deleq \(\qi\)_{\i \in \I}$ is \emph{feasible} if it satisfies the following constraints:
\begin{subequations}\label{eq:feasible_q}
\begin{align}
\sum_{\r \in \R}\qtir&=\lambi D, \quad \forall \ti \in \Ti, \quad \forall \i \in \I, \label{sub:demand}\\
\qtir &\geq 0, \quad \forall \r \in \R, \quad \forall \ti \in \Ti, \quad \forall \i \in \I. \label{sub:non-negative}
\end{align}
\end{subequations}
For a given size vector $\lamb$, let $\Qi(\lamb)$ denote the set of all feasible strategies of population $\i$. From \eqref{sub:demand}-\eqref{sub:non-negative}, we know that the set of feasible strategy profiles $\Q(\lamb) \deleq \prod_{\i \in \I} \Qi(\lamb)$ is a convex polytope.

\subsection{Bayesian Routing Game}\label{sub:game}
\begin{figure}[H]
\centering \setlength{\unitlength}{1cm}  
\begin{picture}(9.5,1.5) 
        \put(0,1){\line(1,0){10}}
            \put(0,1){\makebox(0,0){$\bullet$}}   
        \put(0,0.8){\makebox(0,0){\footnotesize{Nature draws $\s$}}}
        \put(0,0.5){\makebox(0,0){\footnotesize{Population $\i$ receives $\ti$}}}
    \put(5.5,1){\makebox(0,0){$\bullet$}}
     \put(5.5,1.3){\makebox(0,0){{\textit{interim stage}}}}
 \put(5.5,0.8){\makebox(0,0){\footnotesize{Population $\i \in \I$:}}}
 \put(6,0.48){\makebox(0,0){\footnotesize{-obtains belief $\muti$}}}
 \put(5.4,0.1){\makebox(0,0){\footnotesize{-plays strategy $\qi$}}}
      \put(10,0.9){\line(0,1){0.2}}
         \put(10,1){\makebox(0,0){$\bullet$}}
        \put(10,1.3){\makebox(0,0){\textit{ex post stage}}}
        
        \put(10.0,0.8){\makebox(0,0){\footnotesize{Realize costs}}}
\end{picture} 
\caption{Timing of the game $\game$.}
\label{fig:game_timing}
\end{figure}
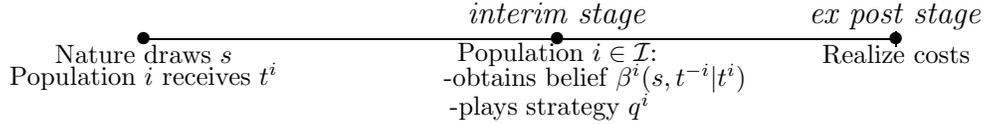

The Bayesian routing game for a fixed size vector $\lambda$ can be defined as \[\game \deleq \left(\I, \S, \T,  \Q(\lamb), \cfunset, \belief\right),\]
where
\begin{itemize}
\item[-] $\I$: Set of populations, $\I = \{1,2, \dots, I\}$
\item[-] $\S$: Set of states with prior distribution $\ps \in \Delta (\S)$ 
\item[-] $\T=\prod_{\i \in \I}\Ti$: Set of population type profiles with element $\t=(\ti)_{\i \in \I} \in \T$
\item[-] $\Q(\lamb)=\prod_{\i \in \I} \Qi(\lamb)$: Set of feasible strategy profiles for a given size vector $\lamb$, with element $\q=(\qi)_{\i \in \I} \in \Q(\lamb)$
\item[-] $\cfunset= \{\cesfun\left(\cdot\right)\}_{\e \in \E, \s \in \S}$: Set of state-dependent edge cost functions
\item[-] $\belief=\left(\belief^{\i}\right)_{\i \in \I}$: $\belief^{\i} \in \Delta(\S \times \Tmi)$ is population $\i$'s belief on state $\s$ and other populations' types $\tmi$
\end{itemize}
All parameters including the common prior $\pi$ are common knowledge, except that populations privately receive signals about the network state from their respective TIS. The game is played as shown in Fig. \ref{fig:game_timing}.

For any $\i \in \I$ and $\ti \in \Ti$, the interim belief of population $\i$ is derived from the common prior:
\begin{equation}\label{interim_belief}
\muti=\frac{\pi(\s, \ti, \tmi)}{\pro(\ti)}, \quad \forall \s \in \S, \quad \forall \tmi \in \Tmi,
\end{equation}
where $\pro(\ti)=\sum_{\s \in \S} \sum_{\tmi \in \Tmi} \pi(\s, \ti, \tmi)$. For a strategy profile $\q  \in \Q(\lamb)$, the induced route flow is denoted $\l \deleq \(\lr(\t)\)_{\r \in \R, \t \in \T}$, where $\lr(\t)$ is the aggregate flow assigned to the route $\r\in \R$ by populations with type profile $\t$, i.e.
\begin{equation}\label{eq:load}
\lr(\t)=\sum_{\i \in \I} \qtir, \quad \forall \r \in \R,  \quad \forall \t \in \T.
\end{equation}
Note that the dependence of $\l$ on $\q$ is implicit and is dropped for notational convenience. 

Again, for the strategy profile $\q \in \Q(\lamb)$, we denote the induced edge load as $w \deleq\(\we(\t)\)_{\e \in \E, \t \in \T}$, where $\we(\t)$ is the aggregate load on the edge $\e$ assigned by populations with type profile $\t$:
\begin{equation}\label{eq:q_w}
\we(\t)=\sum_{\r \ni \e} \sum_{\i \in \I} \qtir \stackrel{\text{\eqref{eq:load}}}{=}\sum_{\r \ni \e} \lr(\t) , \quad \forall \e \in \E, \quad \forall \t \in \T.
\end{equation}
The corresponding cost of edge $\e \in \E$ in state $\s \in \S$ is $\cesfun(\we(\t))$. Then, the cost of route $\r \in \R$ in state $\s \in \S$ can be obtained as: $c_{\r}^{\s}(q(\t))=\sum_{\e \in \r} \cesfun(\we(\t))$. Finally, the expected cost of route $\r$ for population $\i \in \I$ can be expressed as follows:
\begin{align}
\noindent\ecrti  &=\sum_{\s \in \S} \sum_{\tmi \in \Tmi} \sum_{\e \in \r} \muti \cesfun(\we(\ti, \tmi)),
\quad  \forall \r \in \R, \quad \forall \ti \in \Ti, \quad \forall \i \in \I, \label{eq:ecrti}
\end{align}

The equilibrium concept for our game $\game$ is Bayesian Wardrop equilibrium (BWE). A strategy profile $\qwe \in \Q(\lamb)$ is a BWE if for any $\i \in \I$ and any $\ti \in \Ti$: 
\begin{align}\label{eq:BWE_fun}
\forall \r \in \R, \quad \qwetir>0 \quad \Rightarrow \quad \ecrtiwe \leq \ecrptiwe, \quad \forall \r' \in \R.
\end{align}
That is, in a BWE, travelers in population $\i$ with type $\ti$ only take routes that have the smallest expected cost based on their interim belief $\muti$. 

We define the \emph{equilibrium population cost}, denoted $\cilamb$, as the expected cost incurred by a traveler of a given population across all types and network states in equilibrium: $\cilamb \deleq \frac{1}{\lambi \D} \sum_{\ti \in \Ti} \pro(\ti) \sum_{\r \in \R} \ecrtiwe \qwetir$. In fact, from \eqref{eq:BWE_fun}, we can write: 
\begin{align}\label{eq:population_cost_def}
\cilamb \stackrel{\eqref{eq:BWE_fun}}{=} \frac{1}{\lambi \D} \sum_{\ti \in \Ti} \pro(\ti) \(\sum_{\r \in \R} \qwetir\) \min_{\r \in \R} \ecrtiwe \stackrel{\eqref{sub:demand}}{=} \sum_{\ti \in \Ti} \pro(\ti) \min_{\r \in \R} \ecrtiwe.
\end{align}
Note that $\lambi=0$ is a degenerate case for population $\i$ as its size approaches 0. In this case, the cost $\cilamb$ can be viewed as the expected cost faced by an individual (non-atomic) traveler who subscribes to the TIS $\i$.   

%

\section{Equilibrium Characterization}\label{Sec:potential_game}
In this section, we show that the game $\game$ is a weighted potential game. This property enables us to express the sets of equilibrium strategy profiles and route flows as optimal solution sets of certain convex optimization problems. 
\subsection{Equilibrium Strategy Profiles}\label{sub:weighted_potential}
Following \cite{sandholm2001potential}, the game $\game$ is a weighted potential game if there exists a continuously differentiable function $\potential: \Q(\lamb) \rightarrow \mathbb{R}$ and a set of positive, type-specific weights $\{\gamma(\ti)\}_{\ti \in \Ti, \i \in \I}$ such that:
\begin{equation}\label{potential_def}
\frac{\partial \potential(q(\t))}{\partial \qtir}=\gamma(\ti) \ecrti, \quad \forall \r \in \R, \quad \forall \ti \in \Ti, \quad \forall \i \in \I.
\end{equation}
We show that our game $\game$ is a weighted potential game. 
\begin{lemma}\label{potential_lemma}
Game $\game$ is a weighted potential game with the potential function $\potential$ as follows:
\begin{equation}\label{eq:potential_q}
\potential\(q\) \deleq \sum_{\s \in \S} \sum_{\e \in \E} \sum_{\t \in \T} \pi \(\s, \t\) \int_{0}^{\sum_{\r \ni \e} \sum_{\i \in \I} \qtir} \cesfun(z) dz,
\end{equation}
and the positive type-specific weight is $\gamma(\ti) = \pro(\ti)$ for any $\ti \in \Ti$ and any $\i \in \I$. 
\end{lemma}

Using \eqref{eq:load} and \eqref{eq:q_w}, $\potential$ can be equivalently expressed as a function of the route flow $\l$ or the edge load $\w$ induced by a strategy profile $\q \in \Q(\lamb)$: 
\begin{align}
\lpotential(\l) &\deleq \sum_{\s \in \S} \sum_{\e \in \E} \sum_{\t \in \T} \pi \(\s, \t\) \int_{0}^{\sum_{\r \ni \e}\lr(\t)} \cesfun(z) dz \label{eq:potential_l}\\
\wpotential(\w) &\deleq \sum_{\s \in \S} \sum_{\e \in \E} \sum_{\t \in \T} \pi \(\s, \t\) \int_{0}^{\we(\t)} \cesfun(z) dz. \label{eq:potential_w}
\end{align}
Thus, for any feasible strategy profile $\q \in \Q(\lamb)$, we can write $\potential (\q) \equiv  \lpotential(\l) \equiv \wpotential(\w)$, where $\l$ and $\w$ are the route flow and edge loads induced by the strategy profile $\q$. In addition, $\wpotential(\w)$ satisfies the following property:
\begin{lemma}\label{lemma:potential_convex}
The function $\wpotential(\w)$ is twice continuously differentiable and strictly convex in $w$.
\end{lemma}


Our first result provides a characterization of the set of equilibrium strategy profiles:
\begin{theorem}\label{q_opt}
A strategy profile $\q \in \Q(\lamb)$ is a BWE if and only if it is an optimal solution of the following convex optimization problem:
\begin{equation}\tag{OPT-$\Q$}\label{eq:potential_opt}
\min \quad \potential(\q), \quad s.t. \quad \q \in \Q(\lamb),
\end{equation}
where $\Q(\lamb)$ is the set of feasible strategy profiles. The equilibrium edge load vector $\wwe(\lamb)$ is unique.
\end{theorem}

The existence of BWE follows directly from Theorem \ref{q_opt}. For any size vector $\lamb$, we denote the set of BWE for the game $\game$ as $\Qwe(\lamb)$. Importantly, since the equilibrium edge load $\wwelamb$ is unique, the equilibrium population cost $\cilamb$ for each population $\i \in \I$ in \eqref{eq:population_cost_def} must also be unique for any $\lamb$. Thus, the equilibria of $\game$ can be viewed as \emph{essentially unique}. We denote the optimal value of \eqref{eq:potential_opt}, i.e. the value of the weighted potential function $\potential(\q)$ in equilibrium as $\Plamb$. 

The Lagrangian of \eqref{eq:potential_opt} that we use in proving Theorem \ref{q_opt} is given as follows:
\begin{equation}\label{lagrangian}
\lag(q, \mu, \nu, \lamb)=\potential(q)+\sum_{\i \in \I} \sum_{\ti \in \Ti}\alti\(\lamb^i D - \sum_{\r \in \R} \qtir\) -\sum_{\r \in \R}\sum_{\i \in \I} \sum_{\ti \in \Ti}\betir \qtir,
\end{equation}
where $\mu=(\alti)_{\ti \in \Ti, \i \in \I}$ and $\nu=(\betir)_{\r \in \R, \ti \in \Ti, \i \in \I}$ are Lagrange multipliers associated with the constraints \eqref{sub:demand} and \eqref{sub:non-negative}, respectively. The next lemma shows that for any BWE $\qwe \in \Qwe(\lamb)$, the optimal Lagrange multipliers $\mu^{*}$ and $\nu^{*}$ in \eqref{lagrangian} associated with $\qwe$ are unique. 
\begin{lemma}\label{unique_lag}
The Lagrange multipliers $\mu^{*}$ and $\nu^{*}$ at the optimum of \eqref{eq:potential_opt} are unique:
\begin{subequations}
\begin{align}
\mu^{\ti *}&=\min_{\r \in \R}\pro(\ti) \ecrtiwe, \quad \forall \ti \in \Ti, \quad \forall \i \in \I \label{define_al}\\ 
\nu_\r^{\ti *}&= \pro(\ti) \ecrtiwe-\mu^{\ti *}, \quad \forall \r \in \R, \quad \forall \ti \in \Ti, \quad \forall \i \in \I. \label{define_beta}
\end{align}
\end{subequations}
\end{lemma}
This result follows from the fact that \eqref{eq:potential_opt} satisfies the \emph{Linear Independence Constraint Qualification (LICQ)} condition (\cite{wachsmuth2013licq}), which ensures the uniqueness of Lagrange multipliers at the optimum of \eqref{eq:potential_opt}; see Lemma \ref{licq_lemma} for the statement of LICQ condition. 

The value of $\mu^{\ti*}$ relates the expected route costs for each type $\ti$ in equilibrium with the sensitivity analysis of the Lagrangian with respect to population sizes at the optimum of \eqref{eq:potential_opt}. We will use this result in Section \ref{equilibrium_regime} for studying the relative ordering of equilibrium population costs.

\subsection{Equilibrium Route Flows}
Our main question of interest is how the set of BWE $\Qwe(\lamb)$, i.e. optimal solution set of \eqref{eq:potential_opt}, and more importantly, the equilibrium edge load $\wwe(\lamb)$, change with the perturbations in the size vector $\lamb$. However, characterizing the effect of $\lamb$ directly from \eqref{eq:potential_opt} is not so straightforward. Recall that in the simple routing game in Section \ref{motivate}, the effects of perturbations in $\lambda$ on the equilibrium route flow are relatively easier to describe in comparison to the effects on the set of equilibrium strategy profile --- the equilibrium route flow remains fixed in a certain range of $\lamb$, whereas the set of equilibrium strategy profiles do not. 
Thus, our approach involves studying how $\lamb$ affects the set of equilibrium route flows. We show two results in this regard: (i) The set of feasible route flows and the set of feasible strategy profiles that induces a particular route flow can be both expressed as polytopes (Proposition \ref{Lprime}); (ii) The set of equilibrium route flows is the optimal solution set of a convex optimization problem (Proposition \ref{l_opt}). These results enable us to evaluate how the equilibrium edge load and population costs change with perturbations in $\lamb$. 

Let us start by introducing the set of route flows 
\[\F(\lamb) \deleq \{\l \in \mathbb{R}^{|\R| \times |\T|} \left \vert \text{$\l$ satisfies \eqref{sub:balance}-\eqref{sub:popu_i}} \right.\},\] 
where the constraints are given by:
\begin{footnotesize}
\begin{subequations}\label{eq:Lprime}
\begin{align}
\lr(\ti, \tmi)- \lr(\titil, \tmi)&= \lr(\ti, \tmitil)-\lr(\titil, \tmitil), \quad \forall \r \in \R, \text{ } \forall \ti,  \titil \in \Ti, \text{and } \forall \tmi, \tmitil \in \Tmi, \forall \i \in \I,\label{sub:balance}\\
\sum_{\r \in \R} \lr(\t)&=D, \quad \forall \t \in \T, \label{sub:ldemand}\\
\lr(\t) &\geq 0, \quad \forall \r \in \R, \quad \forall \t \in \T, \label{sub:l_positive}\\
\D-\sum_{\r \in \R}\min_{\ti \in \Ti}\lr(\ti, \tmi) &\leq \lambi \D,  \quad  \forall \tmi \in \Tmi, \quad \forall \i \in \I. \label{sub:popu_i}
\end{align} 
\end{subequations}
\end{footnotesize}
The constraints \eqref{sub:balance}-\eqref{sub:l_positive} do not depend on the size vector $\lamb$ and can be understood as follows: \eqref{sub:balance} captures the fact that the change in the flow through any route resulting from change in the type of population $\i \in \I$ does not depend on the particular types of the remaining populations; \eqref{sub:ldemand} ensures that all the demand $\D$ is routed through the network; and \eqref{sub:l_positive} guarantees that the demand assigned to any route is nonnegative. 

On the other hand, the constraints in \eqref{sub:popu_i} depend on the size vector $\lamb$, wherein the size of each population $\i \in \I$, $\lambi$, appears linearly in the constraint corresponding to that population. To further interpret \eqref{sub:popu_i}, we define the ``impact of information'' for any given population as the maximum extent to which the signal received from its TIS can influence the routing behavior of travelers within the population. Specifically, for any strategy profile $\q \in \Q(\lamb)$ and population $\i \in \I$, we define the \emph{impact of information} on population $\i \in \I$ as follows:
\begin{align}\label{eq:mdi}
\mdi(\q) \deleq \lamb^i \D -\sum_{\r \in \R}\min_{\ti \in \Ti} \qtir.
\end{align}
Using \eqref{sub:demand}, we can re-write \eqref{eq:mdi} as: $\mdi(\q)=\sum_{\r \in \R} \max_{\ti \in \Ti}\(q^i_r(\widehat{t}^{i})-q^i_r(\ti)\)$, where $\widehat{t}^{i}$ is an arbitrary type in $\Ti$. That is, for each population $\i \in \I$, $\mdi(\q)$ is the summation over all routes of the maximum difference between the demands assigned to each route $\r$ by the type $\widehat{t}^{i}$ and any other type $\ti \in \Ti$. Using \eqref{eq:load}, we can alternatively express this metric in terms of the flow $\l$ induced by $\q$:
\begin{equation}\label{widehatj}
\begin{split}
\widehat{\md}^{\i}(\l)& \equiv \mdi(\q) \stackrel{\eqref{eq:load}}{=} \sum_{\r \in \R} \max_{\ti \in \Ti}(\lr(\widehat{t}^{i}, \widehat{t}^{-i})-\lr(\ti, \widehat{t}^{-i}))\stackrel{\eqref{sub:ldemand}}{=}\D-\sum_{\r \in \R} \min_{\ti \in \Ti} \lr(\ti, \widehat{t}^{-i}), 
\end{split}
\end{equation}
where $(\widehat{t}^{i}, \widehat{t}^{-i})$ is any type profile in $\T$. Now the constraints \eqref{sub:popu_i} can be equivalently stated as:
\begin{subequations}
\begin{align}
\widehat{\md}^{i}(\l) &\leq \lambi \D, \quad \forall \i \in \I. \tag{IIC} \label{prime:popu_i}
\end{align}
\end{subequations}
These constraints ensure that the impact of signals on any population's strategy is bounded by its size. We will refer to them as information impact constraints (IIC). We use \eqref{prime:popu_i} and \eqref{sub:popu_i} interchangeably, and refer the constraint in \eqref{prime:popu_i} corresponding to population $\i \in \I$ as (\ref{prime:popu_i}$_\i$).  Also, it is easy to see that for each $\i \in \I$,  (\ref{prime:popu_i}$_\i$) can be written as a set of affine inequalities:
\begin{align}\label{affine_popu_i}
\D-\sum_{\r \in \R}\lr(\ti_\r, \tmihat) &\leq \lambi \D, \quad \forall \ti_1 \in \Ti, \dots, \forall \ti_{|\R|} \in \Ti. 
\end{align}
Thus, $\F(\lamb)$ is a convex polytope. The following proposition relates the set of feasible strategy profiles and the induced route flows.

\begin{proposition}\label{Lprime}
The set of feasible route flows is the convex polytope $\F(\lamb)$. Furthermore, for a given route flow $\l \in \F(\lamb)$, any feasible strategy profile $\q\in \Q(\lamb)$ that induces $\l$ can be expressed as:
\begin{align}\label{eq_rep}
\qtir&=\lr(\ti, \tmihat)-\lr(\tihat, \tmihat)+\chi_\r^{\i}, \quad \forall \r \in \R, \quad \forall \ti \in \Ti, \quad \forall \i \in \I,
\end{align}
where $\that=\(\tihat\)_{\i \in \I}$ is any type profile in $\T$, and $\chi=\(\chi_{\r}^{\i}\)_{\r \in \R, \i \in \I}$ satisfies the following constraints:
\begin{subequations}\label{X}
\begin{align}
\sum_{\r \in \R} \chi_\r^\i&=\lambi \D,  \quad \forall \i \in \I, \label{sub:x_sum}\\
\sum_{\i \in \I} \chi_{\r}^{\i}& =\lr( \that), \quad \forall \r \in \R,\label{sub:sum_demand}\\
\chi_\r^\i & \geq \max_{\ti \in \Ti} \(\lr(\tihat, \tmihat)-\lr(\ti, \tmihat) \), \quad  \forall \r \in \R, \quad \forall \i \in \I.\label{sub:x_bound}
\end{align}
\end{subequations}
\end{proposition} 
The proof of this proposition is comprised of three steps: In Step I, we show that any route flow that is induced by a feasible strategy profile must satisfy constraints \eqref{sub:balance}-\eqref{sub:popu_i}. In Step II, we show the converse: for any given $\l$, any feasible strategy profile $\q$ that induces it must be given by \eqref{eq_rep}, where $\chi$ is a vector satisfying \eqref{X}. In Step III, we prove that if $\l$ satisfies \eqref{sub:balance}-\eqref{sub:popu_i}, then we can indeed construct a vector $\chi$ that satisfies \eqref{X}, and the corresponding $\q$ in \eqref{eq_rep} is a feasible strategy profile that induces $\l$. By combining Steps II and III, we can conclude that any $\l$ satisfying \eqref{sub:balance}-\eqref{sub:popu_i} can be induced by at least one feasible strategy profile, and thus is a feasible route flow.

The next proposition provides a characterization of the set of equilibrium route flows, and is analogous to Theorem \ref{q_opt} which characterizes the set of equilibrium strategy profiles.
\begin{proposition}
\label{l_opt}
A feasible route flow $\l \in \F(\lamb)$ is an equilibrium route flow if and only if $\l$ is an optimal solution of the following convex optimization problem:
\begin{equation}\tag{OPT-$\F$}\label{opt_l}
\begin{split}
\min \quad &\lpotential(\l), \quad 
s.t. \quad \l \in \F(\lamb),
\end{split}
\end{equation}
where $\F(\lamb)$ is the set of feasible route flow vectors, which satisfy constraints \eqref{sub:balance} -- \eqref{sub:popu_i}.
\end{proposition}

We denote the set of equilibrium route flows $\lwe$ in the game $\game$ as $\Lwe(\lamb)$. From Theorem \ref{q_opt}, equations \eqref{eq:q_w} and \eqref{eq:potential_q}, we know that for any size vector $\lamb$, and any $\qwe \in \Qwe(\lamb)$, any $\lwe \in \Lwe(\lamb)$, 
\begin{equation}\label{Psi}
\Plamb = \potential(\qwe)=\lpotential(\lwe) =\wpotential(\wwelamb).
\end{equation}

Propositions \ref{Lprime} and \ref{l_opt} form the basis of our analysis of how the perturbations of size vector effects the equilibrium structure and population costs. 
\section{Pairwise Comparison of Populations}\label{equilibrium_regime}
In this section, we first analyze the effects of perturbations in the relative sizes of any two populations on the equilibrium structure. Next, we study how the cost difference between any two populations depends on the population sizes.  

\subsection{Equilibrium Regimes}\label{sub:impact_information}
To study the effects of perturbations in the relative sizes of any two populations, we employ the notion of directional perturbation of size vector $\lamb$. In particular, for any two populations $\i$ and $\j$, we consider the $|\I|$-dimensional direction vector $\zij\deleq(\dots 0 \dots, 1, \dots 0 \dots, -1,$ $ \dots 0 \dots)$ with 1 in the $\i$-th position and $-1$ in the $\j$-th position. When $\lamb$ is perturbed in the direction of $\zij$, the size of population $\i$ (resp. population $j$) increases (resp. decreases), and the sizes of the remaining populations do not change. 

For any size vector $\lamb$ and any two populations $\i$ and $j$, let the vector of the remaining populations' sizes be denoted $\lambmimj \deleq \(\lambk\)_{\k \in \I \setminus \{\i, \j\}}$. The total size of the remaining populations is $|\lambmimj| \deleq \sum_{\k \in \I \setminus \{\i, \j\}} \lambk$. For pairwise comparison, we only consider the case when the sizes of both populations are strictly positive so that $|\lambmimj| < 1$, and the range of the perturbations in the population $\i$'s size is $(0, 1-|\lambmimj|)$. We denote the set of admissible $\lambmimj$ as $\Lambmimj$.

Now consider an optimization problem that is similar to \eqref{opt_l}, except that the two constraints in the \eqref{prime:popu_i} set corresponding to the populations $\i$ and $\j$ are replaced by a single constraint:
\begin{equation}\label{eq:basic_opt_ij}\tag{OPT-$\F^{ij}$}
\min \quad  \lpotential(\l), \quad 
s.t. \quad  \text{\eqref{sub:balance}, \eqref{sub:ldemand}, \eqref{sub:l_positive}, \eqref{prime:popu_i}$\setminus\{\i, \j\}$, \eqref{extra}},
\end{equation}
where the constraints \eqref{prime:popu_i}$\setminus\{\i, \j\}$ indicate that all but (\ref{prime:popu_i}$_\i$) and (\ref{prime:popu_i}$_\j$) from the original set \eqref{prime:popu_i} are included, and the constraint \eqref{extra} is defined as follows: 
\begin{align}\label{extra}\tag{\ref{prime:popu_i}$_{\i\j}$}
\mdl^{\i}(\l)+\mdl^{\j}(\l) \leq \(1-|\lambmimj|\) \D.
\end{align}
The constraint \eqref{extra} ensures that the total impact of information on population $\i$ and $\j$ does not exceed their total demand. We denote the set of optimal solutions for \eqref{eq:basic_opt_ij} as $\Ldagij$. Analogously to Theorem \ref{q_opt}, we can show that any $\ldagij \in \Ldagij$ induces a unique edge load $\wdagij$, which can be obtained by \eqref{eq:q_w}; see Lemma \ref{unique_wdagij}. Then, the optimal solution set of \eqref{eq:basic_opt_ij} can be written as the following polytope:
\begin{equation}\label{drop_i_j}
\begin{split}
\Ldagij = \left\{\l \left \vert
\begin{array}{l}
\text{$\l$ satisfies \eqref{sub:balance}, \eqref{sub:ldemand}, \eqref{sub:l_positive}, \eqref{prime:popu_i}$\setminus\{\i, \j\}$, and \eqref{extra}},\\
\sum_{\r \ni \e} \lr(\t)=\wdagij_\e(\t), \quad \forall \e \in \E, \quad \forall \t \in \T
\end{array}
\right. \right\}.
\end{split}
\end{equation}
Note that both $\Ldagij$ and $\wdagij$ depend on $\lambmimj$ but do not depend on $\lambi$ or $\lambj$. 

Before proceeding further, we need to define two thresholds for the size of one of the two perturbed populations (say, population $\i$): 
\begin{align}\label{pairwise_threshold}
\lambli&\deleq \frac{1}{\D} \min_{\ldagij 
\in \Ldagij}\left\{\widehat{\md}^{\i}(\ldagij)\right\}, \quad 
\lambupi \deleq \frac{1}{\D} \max_{\ldagij
 \in \Ldagij}\left\{\(1-|\lambmimj|\)\D-\widehat{\md}^{\j}(\ldagij)\right\},
\end{align}
where $\widehat{\md}^{\i}(\ldagij)$ and $\widehat{\md}^{\j}(\ldagij)$ are the impact of information metrics for the population $\i$ and $\j$, respectively. We can check that $\lambli$ and $\lambupi$ are admissible thresholds: 
\begin{lemma}\label{lemma:lambl_lambup}
$0\leq \lambli \leq \lambupi \leq 1-|\lambmimj|$.
\end{lemma}
Additionally, \eqref{pairwise_threshold} can be expressed as linear programming problems, see \eqref{linear_program_lambli}-\eqref{linear_program_lambupi}. These two thresholds play a crucial role in our subsequent analysis. 

We are now ready to introduce the equilibrium regimes that are induced by the relative change in the sizes of populations $i$ and $j$ with fixed sizes of other populations $\lambmimj \in \Lambmimj$. These regimes are defined by the following sets: 
\begin{subequations}\label{regime_one_two_three}
\begin{align}
\Reoneij &\deleq \{\(\lambi, \lambj, \lambmimj\) \left \vert \lamb^\i \in (0,  \lambli) \right.\}, \label{def_reg_one}\\
\Retwoij &\deleq \{\(\lambi, \lambj, \lambmimj\) \left \vert \lamb^\i \in 
[\lambli,  \lambupi]\setminus \{0, 1-|\lambmimj|\} \right.\},\label{def_reg_two}\\
\Rethreeij &\deleq \left\{\(\lambi, \lambj, \lambmimj\) \left\vert \lamb^\i \in ( \lambupi, 1-|\lambmimj|) \right. \right\}. \label{def_reg_three}
\end{align} 
\end{subequations} 
We say that the population $\i$ (resp. population $\j$) is a ``minor population'' in regime $\Reoneij$ (resp. regime $\Rethreeij$) because $\lambi < \lambli$ (resp. $\lamb^{j}<1-|\lambmimj|-\lambupi$). Moreover, neither population is minor in regime $\Retwoij$. 
Note that degenerate situations are possible. In particular, if either one or both of the thresholds $\lambli$ and $\lambupi$ take values in the set $\{0, 1-|\lambmimj|\}$, then from the regime definitions \eqref{def_reg_one}-\eqref{def_reg_three}, the number of regimes are reduced to two (for example, in the the simple routing game in Section \ref{motivate}) or even one regime (see Example \ref{single_TIS}). The following theorem describes the properties of equilibrium route flows in the regimes under the directional perturbations in the size vector $\lamb$.
\begin{theorem}\label{l_behavior}
For any two populations $\i, \j \in \I$, and any given $\lambmimj \in \Lambmimj$, the set of equilibrium route flows $\Lwelamb$ when $\lamb$ is in regime $\Reoneij$ or regime $\Rethreeij$ can be expressed as follows:
\begin{align}\label{eq:l_behavior}
\Lwelamb=\left\{ \argmin \lpotential(\l) \left \vert
\begin{array}{lll}
s.t. &\eqref{sub:balance}, \eqref{sub:ldemand}, \eqref{sub:l_positive}, \eqref{extra} \text{ and \eqref{prime:popu_i}$\setminus \{\j\}$} & \text{ if } \lamb \in \Reoneij \\
s.t. &\eqref{sub:balance}, \eqref{sub:ldemand}, \eqref{sub:l_positive}, \eqref{extra} \text{ and \eqref{prime:popu_i}$\setminus \{\i\}$} & \text{ if } \lamb \in \Rethreeij
\end{array}
\right.
\right\}
\end{align}
In regime $\Reoneij$ (resp. regime $\Rethreeij$), the constraint (\ref{prime:popu_i}$_\i$) (resp. (\ref{prime:popu_i}$_\j$)) is tight in equilibrium. Additionally, in regime $\Retwoij$, we have $\Lwelamb \subseteq \Ldagij$.
\end{theorem}

Essentially, this result is based on how the impact of information on each perturbed population compares with its size; i.e. whether or not the constraint (\ref{prime:popu_i}$_i$) (resp. (\ref{prime:popu_i}$_j$)) for the population $\i$ (resp. populatiom $\j$) is tight in equilibrium. In the first side regime $\Reoneij$, the constraint (\ref{prime:popu_i}$_i$) is tight at optimum of \eqref{opt_l}. This implies that the impact of information extends to the entire demand of the minor population $\i$. In fact, the threshold $\lambli$ is the largest size of population $\i$ for which the impact of information on itself is fully attained. We can argue similarly for the other side regime $\Rethreeij$, where population $\j$ is the minor population; i.e. (\ref{prime:popu_i}$_\j$) is tight at optimum of \eqref{opt_l} and ($1-|\lambmimj|-\lambupi$) is the largest size of population $\j$ such that the impact of information on itself is fully attained.

In contrast to the two side regimes, in the middle regime $\Retwoij$, the sizes of both populations $\i$ and $\j$ are above the threshold sizes $\lambli$ and $(1-|\lambmimj|-\lambupi)$, respectively. We can replace the constraints (\ref{prime:popu_i}$_\i$) and (\ref{prime:popu_i}$_\j$) in the optimization problem \eqref{opt_l} by \eqref{extra} without changing its optimal value, i.e. the optimal value of \eqref{eq:basic_opt_ij} is equal to $\Plamb$. However, since the set $\Ldagij$ (as defined in \eqref{drop_i_j}) contains all route flows that attain the optimal value $\Plamb$ but may not necessarily satisfy the constraints (\ref{prime:popu_i}$_\i$) and (\ref{prime:popu_i}$_\j$), the equilibrium route flow set $\Lwe(\lamb)$ must be a subset of $\Ldagij$. In this regime, the impact of information on neither population is fully attained.

A specialized result derived from Theorem \ref{l_behavior} and Proposition \ref{Lprime} is that in routing games with two heterogeneously informed populations and a parallel-route network, the equilibrium strategy profile in the regimes $\Reone^{12}$ and $\Rethree^{12}$ is unique, see Corollary \ref{cor:strategy}.

Thanks to Theorem \ref{l_behavior}, we can analyze the monotonicity of the value of potential function at equilibrium $\Plamb$ with respect to perturbations of $\lamb$ in the direction $\zij$. 
\begin{proposition}\label{bathtub}
For any two populations $\i, \j \in \I$, and any given $\lambmimj \in \Lambmimj$, under directional perturbations of $\lamb$ along the direction $\zij$, the function $\Plamb$ monotonically decreases in regime $\Reoneij$, does not change in $\Retwoij$, and monotonically increases in $\Rethreeij$. 
Furthermore, the equilibrium edge load vector $\wwe(\lamb)= \wdagij$ if and only if $\lamb \in \Retwoij$. 
\end{proposition}
Following Theorem \ref{l_behavior}, in the side regime $\Reoneij$ (resp. $\Rethreeij$), the set of route flows which satisfy the constraints of the optimization problem in \eqref{eq:l_behavior} increases (resp. decreases) as $\lamb$ is perturbed in the direction $\zij$. Thus, the value of the potential function in equilibrium, $\Plamb$, is non-increasing (resp. non-decreasing) in the direction $\zij$. In fact, since the constraint (\ref{prime:popu_i}$_\i$) (resp.  (\ref{prime:popu_i}$_\j$)) is tight in equilibrium, one can argue that $\Plamb$ strictly decreases (resp. increases) in the direction $\zij$. In contrast, in the middle regime $\Retwoij$, since $\Lwe(\lamb) \subseteq \Ldagij$, we can conclude that $\wwe(\lamb)=\wdagij$. Therefore, $\Plamb=\wpotential(\wdagij)$, which does not change when $\lamb$ is perturbed in the direction $\zij$.

The necessary and sufficient condition for the invariance of $\wwe(\lamb)$ under relative perturbations in the sizes of any two populations in Proposition \ref{bathtub} is a direct consequence of the monotonicity of $\Plamb$ and the uniqueness of $\wwe(\lamb)$. This result is useful in determining the relative ordering of population costs in equilibrium, as discussed next.

%

\subsection{Relative Value of Information}
\label{value_of_information}
We now study the difference between the equilibrium costs of any two populations under perturbations in their relative sizes. For any two populations $\i, \j \in \I$ and size vector $\lamb$, we define the \emph{relative value of information}, denoted $\rvlambij$, as the expected travel cost saving that a traveler in population $\i$ enjoys over a traveler in population $\j$, i.e. $\rvlambij \deleq \cjlamb-\cilamb$. Equivalently, $\rvlambij$ is the expected reduction in the cost faced by an individual traveler when her subscription unilaterally changes from TIS $\i$ to TIS $\j$, while the TIS subscriptions of all other travelers remain unchanged. We say TIS $\i$ is relatively more valuable (resp. less valuable) than TIS $\j$ if $\rvlambij>0$ (resp. $\rvlambij<0$). Similarly, if $\rvlambij=0$, TIS $\i$ is said to be as valuable as TIS $\j$.  

It turns out that, for any given size vector $\lamb$, $\rvlambij$ is closely related to the sensitivity of $\Plamb$ (i.e. the value of the potential function in equilibrium) with respect to the perturbation in the relative sizes of populations $\i$ and $\j$. 
\begin{lemma}\label{lemma:directional_derivative_psi}
The value of the weighted potential function in equilibrium, $\Plamb$ as defined in \eqref{Psi}, is convex and directionally differentiable in $\lamb$. For any $\i, \j \in \I$, 
\begin{align}\label{link_rv_derivative}
\rvlambij = -\frac{1}{\D} \Dplambij,
\end{align}
where $\Dplambij\deleq \lim_{\epsilon \to 0^{+}} \frac{\Psi(\lamb+\epsilon \zij)-\Plamb}{\epsilon}$ is the derivative of $\Plamb$ in the direction $\zij$. 
\end{lemma}
\emph{Proof of Lemma \ref{lemma:directional_derivative_psi}}
The proof involves applying the results on sensitivity analysis of convex optimization problems, as summarized in Lemmas \ref{continuous_phi} and \ref{directional}; for a detailed background on these technical results, we refer the reader to \cite{fiacco2009sensitivity}, \cite{fiacco1986convexity}, and \cite{rockafellar1984directional}. Since in \eqref{eq:potential_opt}, the weighted potential function $\potential(\q)$ is convex in $\q$, and the constraints \eqref{sub:demand}-\eqref{sub:non-negative} are affine in $\q$ and $\lamb$, from Lemma \ref{continuous_phi}, we know that the optimal value of the potential function $\Plamb$ is convex in $\lamb$. 

Next, we can check that \eqref{eq:potential_opt} satisfies the following conditions: (1) The potential function $\potential(\q)$ is continuously differentiable in $\q$, and constraints \eqref{sub:demand}-\eqref{sub:non-negative} are linear in $\q$ and $\lamb$; (2) The optimal solution set $\Qwe(\lamb)$ is non-empty and bounded (Theorem \ref{q_opt}). The Lagrange multipliers at the optimum of \eqref{eq:potential_opt} are unique, and bounded (Lemma \ref{unique_lag}). Therefore, from Lemma \ref{directional}, we know that $\Plamb$ is differentiable in the direction $\zij$, and $\Dplambij$ can be expressed as:
\begin{footnotesize}
\begin{align*}
\Dplambij&=\min_{\qwe \in \Qwe(\lamb)} \max_{\substack{
(\mu^{*}, \nu^{*})\\
\in \(M(\qwe), N(\qwe)\)}} \nabla_{\lamb} \lag(\qwe, \mu^{*}, \nu^{*}, \lamb) z^{\i \j}\\
&\stackrel{\text{\eqref{lagrangian}}}{=}\min_{\qwe \in \Qwe(\lamb)} \max_{\substack{
(\mu^{*}, \nu^{*})\\
\in \(M(\qwe), N(\qwe)\)}}\(\sum_{\ti \in \Ti} \mu^{*\ti}-\sum_{\tj \in \T^\j} \mu^{*\tj}\) \D,
\end{align*}
\end{footnotesize}where $M(\qwe)$ (resp. $N(\qwe)$) is the set of optimal Lagrange multipliers $\mu^{*}$ (resp. $\nu^{*}$) associated with the equilibrium strategy $\qwe$. From Lemma \ref{unique_lag}, since both $\mu^{*}$ and $\nu^{*}$ are unique in equilibrium, $\Dplambij$ can be simplified:
\begin{footnotesize}
\begin{align*}
\Dplambij&=\(\sum_{\ti \in \Ti} \mu^{*\ti}-\sum_{\tj \in \T^\j} \mu^{*\tj}\) \D\\
&\stackrel{\eqref{define_al}}{=}\(\sum_{\ti \in \Ti} \min_{\r \in \R}\pro(\ti) \ecrtiwe -\sum_{\tj \in \T^\j} \min_{\r \in \R}\pro(\tj) \ecrtjwe \) \D\notag\\
&\stackrel{\eqref{eq:population_cost_def}}{=} \(\cilamb-\cjlamb\) \D=-\rvlambij\cdot \D.\hskip0.5\textwidth  
\end{align*}
\end{footnotesize}
\QEDA

Our next theorem provides the qualitative structure of relative value of information in the three regimes \eqref{def_reg_one}-\eqref{def_reg_three}.
\begin{theorem}\label{prop:relative_value}
For any two populations $\i, \j \in \I$, and any $\lambmimj \in \Lambmimj$, the relative value of information $\rvlambij$ is positive in regime $\Reoneij$, zero in regime $\Retwoij$, and negative in regime $\Rethreeij$. 

Furthermore, $\rvlambij$ is non-increasing in the direction $\zij$.
\end{theorem}
\emph{Proof of Theorem \ref{prop:relative_value}.}
First, we know from Proposition \ref{bathtub} that in direction $z^{ij}$, $\Plamb$ decreases in regime $\Reoneij$, does not change in regime $\Retwoij$ and increases in regime $\Rethreeij$. Following Lemma \ref{lemma:directional_derivative_psi}, we directly obtain that $\rvlambij>0$ in $\Reoneij$, $\rvlambij=0$ in $\Retwoij$, and $\rvlambij<0$ in $\Rethreeij$.

Next, from Lemma \ref{lemma:directional_derivative_psi}, we know that $\Plamb$ is convex in $\lamb$. Hence, for any $\i, \j \in \I$, the directional derivative $\Dplambij$ is non-decreasing in $\zij$. From \eqref{link_rv_derivative}, $\rvlambij$ is non-increasing in $\zij$.
\QEDA

Theorem \ref{prop:relative_value} shows that one population has advantage over another population if and only if it is the minor population of the two. For the two side regimes, information impacts the entire demand of the minor population. As a result, in equilibrium, the travelers in the minor population do not choose the routes with a high expected cost based on the signal they receive from their TIS;  however, the travelers in the other population may still choose these routes. On the other hand, in the middle regime, neither population has an advantage over the other one because the information only partially impacts each population's demand. Consequently, both populations route their demand in a manner such that they face identical cost in equilibrium.

Additionally, the travel cost saving that population $\i$ travelers enjoy over the population $\j$ is the highest when population $\i$ has few travelers. Intuitively, in each side regime, the travelers in the non-minor population face a higher congestion externality relative to the travelers in the minor population, because all travelers within a population are routed according to the same strategy.  Naturally, the difference in the equilibrium costs due to the imbalance in congestion externality decreases as the size of the minor population increases, and reduces to zero in the middle regime. 


Furthermore, given any two populations $\i, \j \in \I$ and the sizes of all other populations $\lambmimj$ being fixed, Theorem \ref{prop:relative_value} provides a computational approach to compare the equilibrium costs of populations $\i$ and $\j$ for the full range of $\lambi \in (0, 1-|\lambmimj|)$ without explicit computation of BWE or equilibrium route flows for each $\lambi$. This approach can be summarized as follows: (i) Solve \eqref{eq:basic_opt_ij} to obtain an optimal solution $\ldagij$; (ii) Compute $\wdagij$ by plugging $\ldagij$ into \eqref{eq:q_w}; (iii) Obtain $\lambli$ and $\lambupi$ by solving \eqref{pairwise_threshold}; and (iv) Find the relative ordering of equilibrium costs of population $\i$ and $\j$ by checking which of the three possible regimes the size vector $\lamb$ belongs to.

Finally, we can specialize Theorem \ref{prop:relative_value} to analyze situations when a population does not have an access to a TIS, or chooses not to use it at all. Formally, we say that a population $\j \in \I$ is \emph{uninformed} if its type is independent with the network state and other populations' types, i.e. $\pro(\tj|\s, \t^{-j})=\pro(\tj)$ for any $\t^{-j} \in \T^{-j}$, any $\tj \in \T^\j$, and any $\s \in \S$. Following \eqref{interim_belief}, the uninformed population $\j$'s interim belief can be written as follows:
\begin{align}
\mutj&\stackrel{\eqref{interim_belief}}{=}\frac{\pi(\s, \tj, \t^{-j})}{Pr(\tj)}=\frac{Pr(\tj|\s, \t^{-j}) \cdot Pr(\s, \t^{-j})}{Pr(\tj)}=Pr(\s, \t^{-j})=\sum_{\tj \in \T^\j} \pi(\s, \t^{-j}, \tj), 
\label{uninformed_mu}
\end{align} 
That is, the interim belief $\mutj$ is identical for any signal $\tj \in \Tj$ received by population $\j$, and is equal to the marginal probability of $\(\s, \t^{-\j}\)$ calculated from the common prior $\pi$. Therefore, the uninformed population has no further information besides the common knowledge. We show that the equilibrium cost of the uninformed travelers is no less than the cost of any other population.
\begin{proposition}\label{no_info}
Consider the game $\game$ in which population $\j$ is uninformed. Then, for any size vector $\lamb$, the equilibrium cost of population $\j$'s travelers $\cjlamb \geq \cilamb$, 
where the population $\i$ is any other population (i.e. $\i \in \I \setminus \{\j\}$).
\end{proposition} 
Indeed, if population $\j$ is uninformed, we can argue that its equilibrium routing strategy $q^{\j*}(\tj)$ must be identical for any $\tj \in \T^{j}$. Consequently, from \eqref{eq:mdi}, the impact of information metric for the population $\j$ is $\widehat{J}^\j(\q^{\j*})=0$, and perturbing the relative sizes of population $\j$ and any other population $\i \in \I \setminus \{\j\}$ never results in a regime in which population $\j$ is the minor population. Applying Theorem \ref{prop:relative_value}, we can conclude that the equilibrium cost of population $\j$ cannot be less than that of any other population. 

We illustrate the results on equilibrium structure and relative value of information in the following two examples:
\begin{example}\label{benchmark_example}
We consider a game with two populations on two parallel routes ($\rone$ and $\rtwo$) with following parameters: $\ps(a)=0.2$, $\D=10$, $c_1^{\n}(\l_1)=\l_1+15$, $c_1^{\a}(\l_1)=3\l_1+15$, $c_2 (\l_2)=2\l_2+20$. Types $\tone$ and $\ttwo$ are independent conditional on the state, i.e. $\pro(\tone, \ttwo|\s)=\pro(\tone|\s) \cdot \pro(\ttwo|\s)$. Population 1 has 0.8 chance of getting accurate information of the state, and population 2 has 0.6 chance, i.e. $\pro(\tone=\s|\s)=0.8$, and $\pro(\ttwo=\s|\s)=0.6$. The value of the potential function in equilibrium, equilibrium route flows and population costs are shown in Fig. \ref{fig:8_6_load_potential_cost}.

In this example, population 1 travelers receive more accurate state information than population 2 travelers. However, population 1 faces a higher cost than population 2 when its size is sufficiently large, i.e., when $\lamb$ is in regime $\Rethree^{12}$; see Figure \ref{fig:ind_both_inf}. This is due to the fact that in regime $\Rethree^{12}$, the population 1's advantage of receiving more accurate information is dominated by the congestion externality it faces due to its relatively large size, in comparison to population 2.
\begin{figure}[H]
\centering
\begin{subfigure}[b]{0.32\textwidth}
       \includegraphics[width=\textwidth]{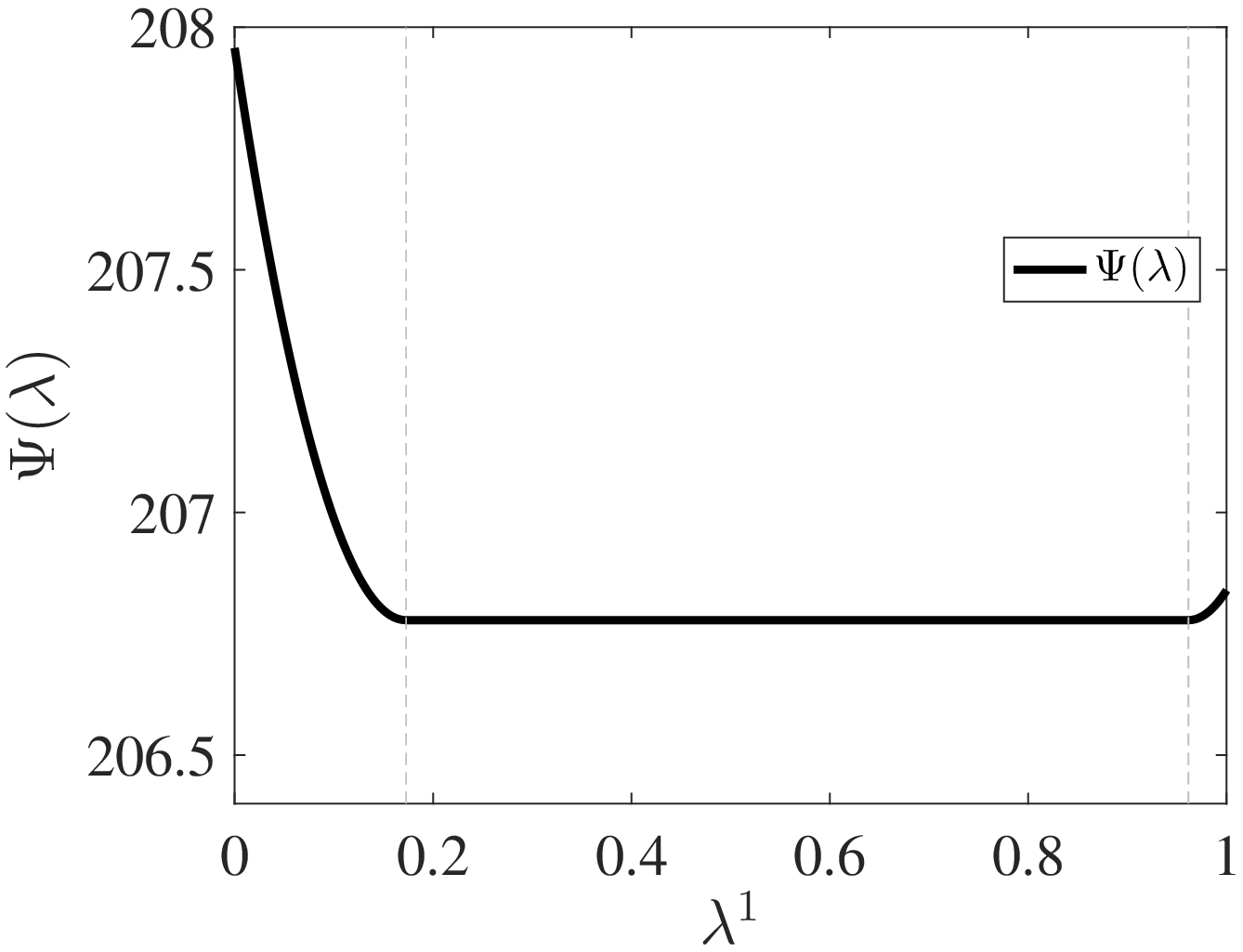}
\caption{}
    \label{fig:pot_80_60_18_20_2}
    \end{subfigure}
~ \begin{subfigure}[b]{0.32\textwidth}
\includegraphics[width=\textwidth]{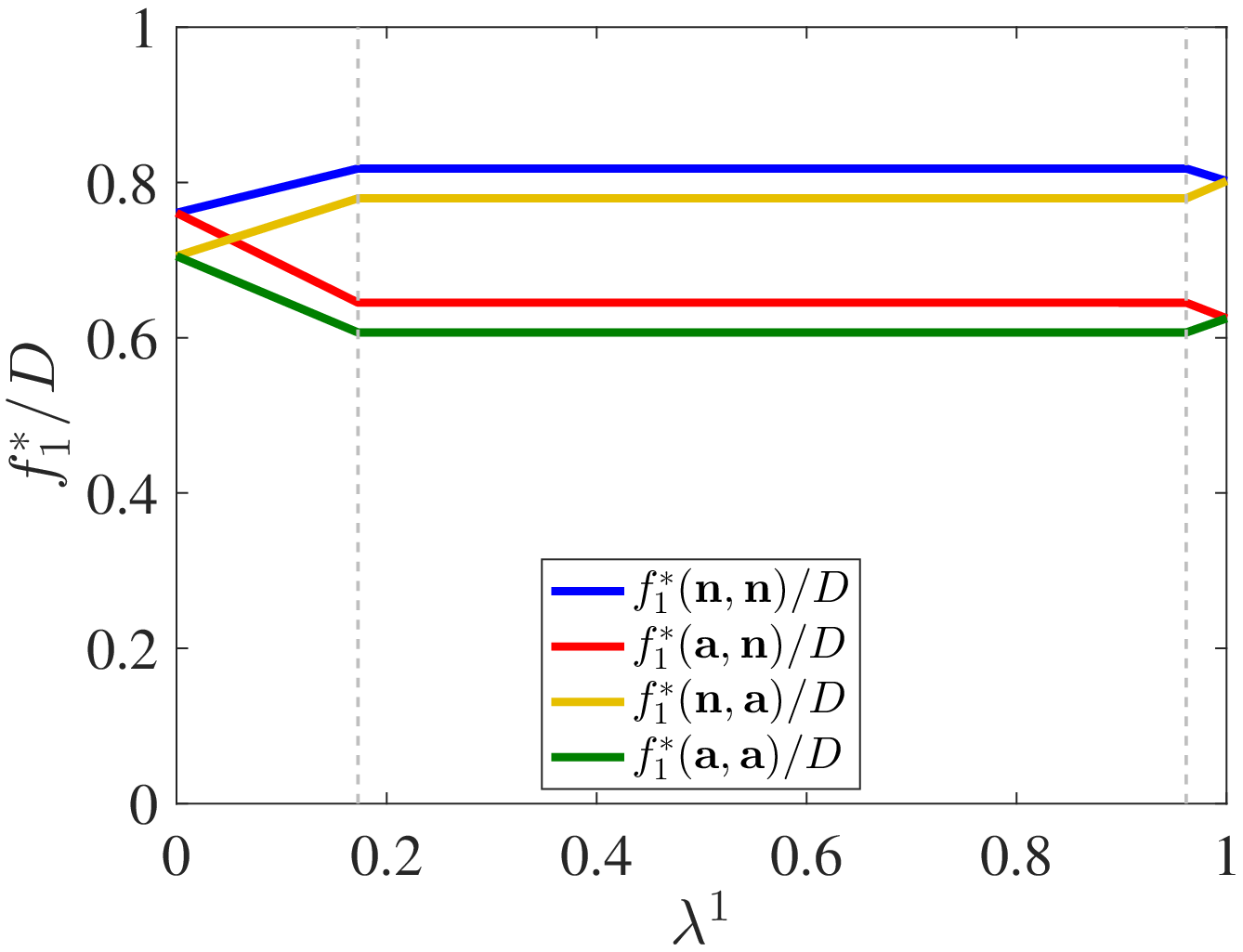}
\caption{}
\label{fig:load_H_80_L_60_D_10_P_2}
\end{subfigure}~
\begin{subfigure}[b]{0.32\textwidth}
       \includegraphics[width=\textwidth]{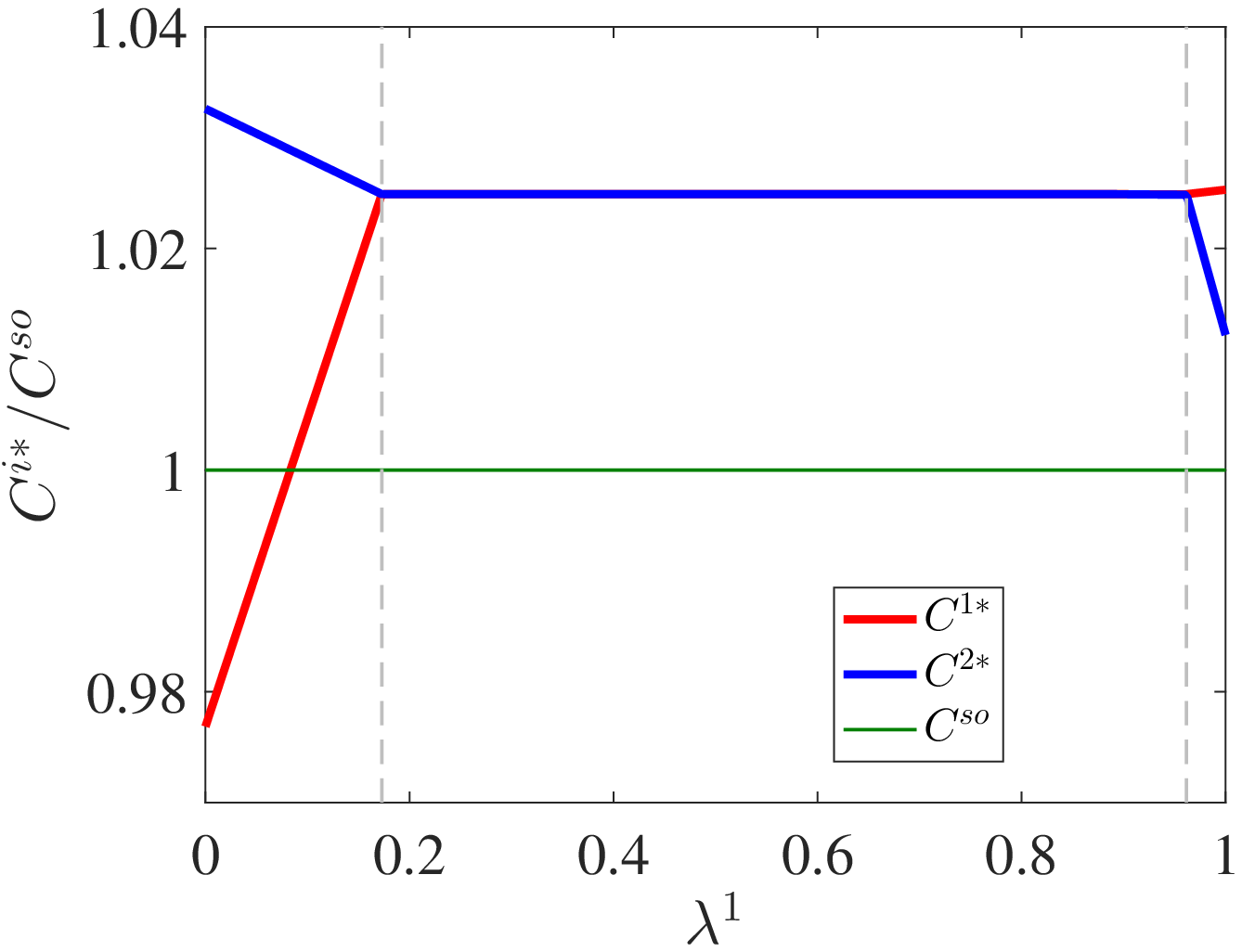}
\caption{}
    \label{fig:ind_both_inf}
    \end{subfigure}
\caption{Effects of varying population sizes for Example \ref{benchmark_example}: (a) Weighted potential function in equilibrium, (b) Equilibrium route flows on $r_1$, and (c) Equilibrium population costs.}
\label{fig:8_6_load_potential_cost} 
\end{figure} 
\end{example}

\begin{example}\label{correlation}
Let us now consider the game with two populations on two parallel routes ($r_1$ and $r_2$) with the same cost functions, prior distribution $\theta$ and total demand $\D$ as that in Example \ref{benchmark_example}. Both populations 1 and 2 have 0.75 chance of getting accurate information about the state, i.e. $\pro(\ti=\s|\s)=0.75$ for any $\i \in \I$ and any $\s \in \S$. In Fig. \ref{fig:compare}, we illustrate the equilibrium population costs in two cases: (i) Types $\tone$ and $\ttwo$ are perfectly correlated, i.e. $\tone=\ttwo$; (ii) Types $\tone$ and $\ttwo$ are independent conditional on the state, i.e. $\pro(\tone, \ttwo|\s)=\pro(\tone|\s) \cdot \pro(\ttwo|\s)$.

This example illustrates how the correlation among received signals (or lack thereof) affects the equilibrium structure. Note that case (i) can be viewed as a single-population game. This is because when $\tone$ and $\ttwo$ are perfectly correlated, there is no information asymmetry among travelers. Thus, $\lamb^1$ has no impact on the equilibrium outcome, and $\lambl=0$, $\lambup=1$ (Fig.~\ref{fig:perfect_correlate}). However, case (ii) is not equivalent to a single-population game. Although both populations have identical chance of getting accurate information about the state, there is information heterogeneity among travelers of the two populations, i.e. travelers in one population do not know the signals received by travelers in the other population, and thus the equilibrium outcome changes with the size $\lamb^1$ (Fig.~\ref{fig:independent}).
\begin{figure}[H]
\centering

\begin{subfigure}[b]{0.32\textwidth}
       \includegraphics[width=\textwidth]{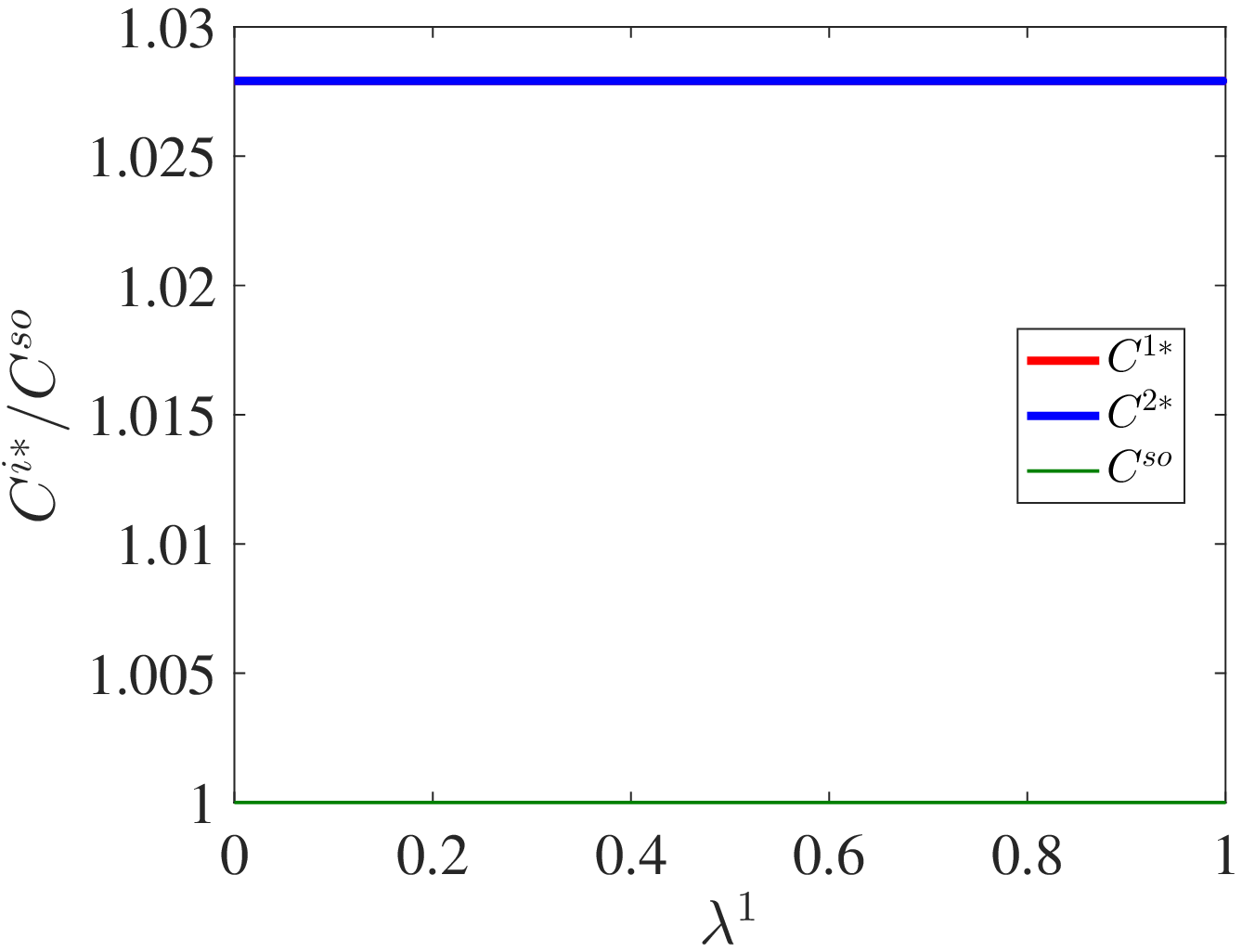}
\caption{}
    \label{fig:perfect_correlate}
    \end{subfigure}
    ~
\begin{subfigure}[b]{0.32\textwidth}
\includegraphics[width=\textwidth]{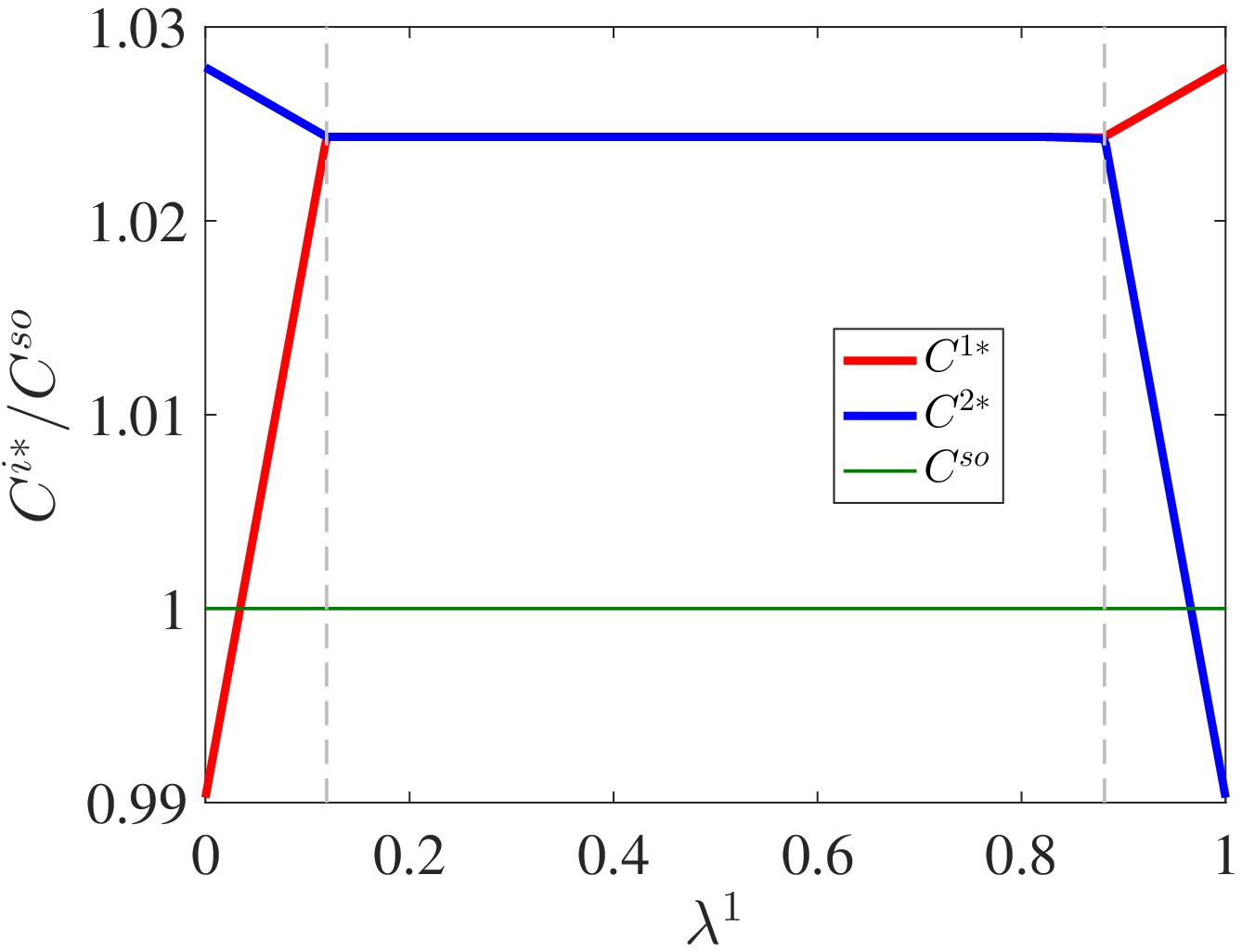}
\caption{}
\label{fig:independent}
\end{subfigure}
\caption{Effects of varying population sizes on equilibrium population costs for Example \ref{correlation}: (a) Perfectly correlated types; (b) Conditional independent types.}
\label{fig:compare} 
\end{figure}
\end{example}

We include two additional examples in the e-companion. Example \ref{non_necessary} shows that the regime $\Rethree^{12}$ can be empty even when population 2 is not an uninformed population. Thus, an uninformed population $\j$ is sufficient but not necessary for $\lambupi=1-|\lambmimj|$. In Example \ref{single_TIS}, we present a situation when only single regime exists in equilibrium.

Our results so far focus on how equilibrium properties and population costs change with the directional perturbation of the size vector $\lamb$. We emphasize that given any $\i, \j \in \I$, the thresholds $\lambli$ and $\lambupi$, as defined in \eqref{pairwise_threshold}, depend on the sizes of the remaining populations $\lambmimj$, and the populations' interim beliefs $(\belief^{\i})_{\i \in \I}$ derived from the common prior $\pi$. Importantly, the qualitative structure of the equilibrium regimes resulting from perturbations in the sizes of any two populations is applicable for any size vector $\lamb$ and any common prior. The main property that drives these results is that the equilibrium regimes only depend on whether or not the impact of information on each population is fully attained.

\section{General Properties of Equilibrium Outcome}\label{sec:general_cost}
In this section, we first extend our approach of pairwise comparison of populations to study how the equilibrium outcome depends on population sizes in general. Then, we analyze the TIS adoption rates in situations where travelers can choose information subscription. 

\subsection{Size-Independence of Edge Load Vector}\label{sub:equilibrium_edge}
Our analysis in Section \ref{equilibrium_regime} showed that if perturbations in the relative sizes of any two populations $\i, \j \in \I$ induce a middle regime $\Retwo^{ij}$, then the equilibrium outcome in this regime is independent of the sizes of the perturbed populations $\i$ and $\j$. A natural question to ask is whether this result can be generalized; i.e., can we find a set of size vectors for which the equilibrium edge load does not depend on the size of \emph{any} population? The answer is affirmative. 

We now explicitly characterize the set of size vectors, denoted $\Lambdag$, for which the edge load is size-independent. 
Since \eqref{opt_l} is a convex optimization problem, and \eqref{prime:popu_i} are the only size-dependent constraints, we can equivalently view $\Lambdag$ as the set of size vectors for which all the IICs can be dropped from \eqref{opt_l} without changing its optimal value. Hence, for any $\lamb \in \Lambdag$, the optimal value of \eqref{opt_l} is identical to that of the following convex optimization problem:
\begin{equation}\label{eq:drop_all}
\begin{split}
\min \quad &\lpotential(\l), \quad
s.t. \quad  \text{\eqref{sub:balance}, \eqref{sub:ldemand} and \eqref{sub:l_positive}}.
\end{split}
\end{equation}
Let us denote the optimal solution set of \eqref{eq:drop_all} as $\Ldag$. Analogous to Theorem \ref{q_opt}, one can argue that any optimal solution $\ldag \in \Ldag$ induces a unique edge load $\wdag$, obtained from \eqref{eq:q_w}. Thus, $\Ldag$ can be written as the convex polytope:
\begin{align}\label{eq:Ldag_polytope}
\Ldag= \left\{\l \left \vert
\begin{array}{l}
\text{$\l$ satisfies \eqref{sub:balance}, \eqref{sub:ldemand}, \eqref{sub:l_positive}, and $\sum_{\r \ni \e} \lr(\t)=\wdag_\e(\t), \quad \forall \e \in \E, \quad \forall \t \in \T$}
\end{array}
\right. \right\}.
\end{align}
Furthermore, since any route flow in the set $\Ldag$ satisfies the constraints \eqref{sub:balance}-\eqref{sub:l_positive} -- but not necessarily \eqref{prime:popu_i} constraints -- and also attains the optimal value of \eqref{opt_l}, we must have that for any $\lamb \in \Lambdag$, $\Lwe(\lamb) \subseteq \Ldag$. Therefore, for each $\lamb \in \Lambdag$, 
there must exist a $\ldag \in \Ldag$ that is an equilibrium route flow, i.e. at least one $\ldag \in \Ldag$ satisfies the \eqref{prime:popu_i} constraints  corresponding to $\lamb$:
\begin{align}\label{eq:intermediate}
\Lambdag\deleq\left\{\lamb \left \vert 
\sum_{\i \in \I} \lambi =1; \quad \lambi \geq 0, \text{ } \forall \i \in \I; \quad
\text{$\exists \ldag \in \Ldag$ s.t. } \widehat{\md}^{i}(\ldag) \leq \lambi \D, \quad \forall \i \in \I
 \right.\right\}
\end{align}
%

The following proposition shows the properties of equilibrium edge load and the value of potential function in the set $\Lambdag$:
\begin{proposition}\label{theorem:intermediate}
The set $\Lambdag$ is convex, and attains the minimum of $\Plamb$, i.e. $\Lambdag=\argmin_{\lamb}\Plamb$. The equilibrium edge load vector $\wwe(\lamb)$ is size-independent, and is equal to $\wdag$ if and only if $\lamb \in \Lambdag$. 
\end{proposition}
This result shows that some of the properties of $\Plamb$ and the change of equilibrium edge load vector under pairwise perturbation (Proposition \ref{bathtub}) also hold for the more general case of perturbation in sizes of multiple populations. 

\subsection{Adoption Rates under Choice of TIS}\label{implication}
Our analysis so far has focused on the equilibrium properties with fixed population sizes. We now extend our results on the relative value of information (Section \ref{equilibrium_regime}) and the size independence of the equilibrium edge load vector (Section \ref{sub:equilibrium_edge}) to analyze travelers' choice of TIS subscription when they can choose to subscribe to any TIS in the set $\I$. 

We model travelers' choice of TIS and the choice of routes as a two-stage game: In the first stage, travelers choose to subscribe to one TIS from the set $\I$. The induced size vector is $\lamb=\(\lambi\)_{\i \in \I}$, where $\lambi$ is the fraction of travelers who choose TIS $\i$. In the second stage, travelers play the Bayesian routing game $\game$. Note that the size vector $\lamb$ here is determined by the travelers' TIS choices in the first stage, as opposed to being a parameter.

In equilibrium, a traveler who chooses TIS $\i$ experiences the expected cost $\cilamb$. The travelers has no ex-ante incentive to unilaterally change her TIS subscription if and only if $\cilamb$ is the lowest across all $\i \in \I$. Therefore, no traveler has the incentive to change her TIS subscription if and only if: 
\begin{align}\label{ssf}
\lambi>0 \quad \Rightarrow \quad \cilamb=\min_{\j \in \I} \cjlamb, \quad \forall \i \in \I. 
\end{align}Such population size vector $\lamb$ can be viewed as the vector of equilibrium adoption rates, one for each TIS. For any size vector that satisfies \eqref{ssf}, all travelers experience identical expected costs, and no traveler has the incentive to change her TIS subscriptions. 

Our next theorem shows that all size vectors $\lamb \in \Lambdag$ are equilibrium TIS adoption rates. 
\begin{theorem}\label{order_population}
The set of equilibrium adoption rates under the choice of TIS is $\Lambdag$. 
\end{theorem}

\emph{Proof of Theorem \ref{order_population}}
Firstly, we show for any $\lamb \in \Lambdag$, all travelers have identical costs in equilibrium. Consider any $\i, \j \in \I$ such that $\lambi>0$ and $\lamb^j>0$, the directional derivative of $\Plamb$ in the direction $\zij$, $\Dplambij$, must be 0. Otherwise, $\Plamb$ strictly decreases in the direction $\zij$ (resp. $z^{ji}$) if $\Dplambij<0$ (resp. if $\Dplambij>0$), which contradicts the fact that $\Lambdag = \argmin_{\lamb} \Plamb$ as in Proposition \ref{theorem:intermediate}. From Lemma \ref{lemma:directional_derivative_psi}, we know that $\cilamb=\cjlamb$. Therefore, any two populations with positive size have identical costs in equilibrium. 

If any $\lamb \in \Lambdag$ satisfies $\lambi>0$ for all $\i \in \I$, then the first step in our proof is sufficient to show that \eqref{ssf} is satiesfied. Otherwise, for any $\lamb \in \Lambdag$, and any degenerate population $\i \in \{\I|\lambi=0\}$, we need to show that $\cilamb \geq \cjlamb$, where $\lambj>0$. Since $\Lambdag = \argmin_{\lamb} \Plamb$, we know that $\Plamb$ must be non-decreasing in the direction $\zij$. Thus, we obtain: $\Dplambij\stackrel{\eqref{link_rv_derivative}}{=}\(\cilamb-\cjlamb\) \D \geq 0$. Thus, $\cilamb \geq \cjlamb$. The first and the second steps together show that any $\lamb \in \Lambdag$ satisfies \eqref{ssf}, and hence is a vector of equilibrium adoption rates.

Finally, we show that for any feasible $\lamb \notin \Lambdag$, \eqref{ssf} is not satisfied. Since $\Lambdag = \argmin_{\lamb} \Plamb$, for any $\lamb \notin \Lambdag$, we can claim that there must exist a direction $\zij$ such that $\Plamb$ decreases in the direction $\zij$, 
$\Dplambij<0$. Otherwise, $\lamb$ is a local minimum of $\Plamb$, and since $\Plamb$ is convex in $\lamb$ (Lemma \ref{lemma:directional_derivative_psi}), $\lamb$ is a global minimum, which contradicts the fact that $\lamb \notin \Lambdag$. For such a direction $\zij$, there are two possible cases: (1) $\lambi >0$ and $\lambj>0$. In this case, from \eqref{link_rv_derivative}, $\cilamb \neq \cjlamb$, and thus travelers do not have identical costs in equilibrium. (2) $\lambi=0$ and $\lambj>0$. In this case, from \eqref{link_rv_derivative}, we must have $\Dplambij=\(\cilamb-\cjlamb\) \D<0$. Therefore, $\cjlamb>\cilamb$, which implies that travelers in population $\j$ has incentive to change subscription to TIS $\i$. To sum up, in either case, $\lamb \notin \Lambdag$ cannot be a vector of equilibrium adoption rates.
\QEDA

\color{black}

Note that the set $\Lambdag$ is not a singleton set in general. Recall from Example \ref{benchmark_example}, the set $\Lambdag$ is the range $\Retwo^{12}$, in which both TIS are chosen. Therefore, the equilibrium adoption rate of each TIS is not unique. However, since $\Lambdag$ is a convex set, the equilibrium adoption rate of each TIS $\i \in \I$ is in a continuous range, denoted $[\lamb^{\i\dagger}_{min}, \lamb^{\i\dagger}_{max}]$, where $\lamb^{\i\dagger}_{min}=\min_{\lamb^{\dagger}\in \Lambdag} \lamb^{\i\dagger}$ (resp. $\lamb^{\i\dagger}_{max}=max_{\lamb^{\dagger} \in \Lambdag} \lamb^{\i\dagger}$) is the minimum (resp. maximum) equilibrium adoption rate. Furthermore, since the set $\Lambdag$ is determined by the heterogeneous information environment created by all TIS, the equilibrium adoption rate of each TIS $\i$ is not only determined by the distribution of its own signal, but is also related to the distribution of other TIS signals, and the possible correlations between signals of different TIS. 

Finally, Theorem \ref{order_population} can be used to assess whether or not a set of TIS can induce a heterogeneous information environment. For any $\lamb^{\dagger} \in \Lambdag$, the support set of $\lamb^{\dagger}$, denoted $\suplamb \deleq \{\i \in \I | \lamb^{\i\dagger} >0\}$, represents the set of TISs chosen by travelers. In particular, if $|\suplamb|=1$, then all travelers choose to subscribe to a single TIS even though multiple TIS are available. Thus, the resulting information environment is homogeneous; see Example \ref{single_TIS}, where $\lamb^1=1$ and $\lamb^2=0$ is the only equilibrium adoption rate. However, if $|\suplamb|>1$, then more than one TIS are chosen, i.e., the heterogeneous information environment is sustained. 
Moreover, if TIS $\i \notin \suplamb$ for any $\lamb^{\dagger} \in \Lambdag$, then this TIS is redundant in that it is not chosen in equilibrium even if it is available to travelers.

%

\section{Concluding Remarks}\label{concluding}
In this article, we study the equilibrium route choices and costs in a heterogeneous information environment, in which each population receives a private signal from their traffic information system (TIS). Each population maintains a belief about the unknown network state and about the signals received by other traveler populations. We focus on analyzing the equilibrium structure under perturbations of population sizes, the relative value of information between any pair of populations, as well as the equilibrium adoption rates when travelers can choose their TIS subscription. 

The main ideas behind our analysis approach are: (i) Identification of qualitatively distinct equilibrium regimes based on whether or not the impact of information is fully attained; (ii) Sensitivity analysis of the weighted potential function in equilibrium with respect to the population size vector; and (iii) Characterization of adoption rates under the choice of TIS. Our approach can be easily extended to games where the edge costs are non-decreasing (rather than strictly increasing) in the edge loads. In particular, such a game still admits a weighted potential function, although now the essential uniqueness only applies to the equilibrium edge costs, rather than the loads. The qualitative properties of equilibrium structure, results about the relative ordering of population costs, and adoption rates can be extended as well. However, the characterization of regime thresholds in this case is more complicated from a computational viewpoint due to the non-uniqueness of edge load vector.

One future research question of interest is to analyze how the travelers' expected cost and TIS adoption rates change when one or more TIS providers make technological changes to their service (for example, improving accuracy levels), or when a new TIS is introduced. Addressing this problem would involve applying our results to evaluate the value of information for each traveler population as well as the adoption rates under the new information environment, and comparing them with that of the current environment. 

Another interesting extension of this work is to study how to design an ``optimal" information structure to regulate traffic flows in a network with uncertain state. In the spirit of Bayesian persuasion viewpoint developed by \cite{kamenica2011bayesian}, and \cite{kolotilin2017persuasion}, in this extension, an information designer (transportation authority) can send a public or private signal that is correlated with the network state to travelers according to a chosen information structure. The signal determines the (heterogeneous) information environment faced by travelers, and hence influences their routing decisions. The approach presented in this article can be useful in analyzing the impact of information design on travelers' routing decisions and costs. 
\section*{Acknowledgement}The authors thank Daron Acemoglu, Tamer Başar, Alexandre Bayen, Sanjeev Goyal, Ali Jadbabaie, Patrick Jaillet, Ramesh Johari, Zaid Khan, Jeffery Liu, Michael Schwarz, Galina Schwartz, Max Shen, David Simchi-Levi, Demos Teneketzis, Pravin Varaiya, Adam Wierman, Muhamet Yildiz for suggestions. We deeply appreciate the feedback of two anonymous reviewers and Associate Editor who reviewed the first version. This work was supported by Google Faculty Research Award, NSF Grant No. 1239054 CPS Frontiers: Foundations Of Resilient CybEr-Physical Systems (FORCES), and NSF CAREER Award CNS 1453126.

\bibliographystyle{plainnat}
\bibliography{library}
\newpage

\begin{appendix}
\setcounter{equation}{0}
\renewcommand{\theequation}{\thesection.\arabic{equation}}
\newtheorem{exam}{Example}[section]
\newtheorem{prop}{Proposition}[section]
\newtheorem{lem}{Lemma}[section]
\newtheorem{coro}{Corollary}[section]
\section{Supplementary Material for Sec. \ref{Sec:potential_game}}\label{appendix_A}
\emph{Proof of Lemma \ref{potential_lemma}.} 
First note that $\potential(\q)$, as defined in \eqref{eq:potential_q}, is a continuous and differentiable function of the strategy profile $\q$. To show that $\potential(\q)$ is a weighted potential function of $\game$, we write the first order derivative of $\potential(q)$ with respect to $\qtir$:
\begin{align}
\frac{\partial \potential (q)}{\partial \qtir}&=\sum_{\s \in \S} \sum_{\tmi \in \Tmi} \pi(\s, \ti, \tmi) \sum_{\e \in \r}\cesfun\(\we(\ti, \tmi)\) \notag\\
&\stackrel{\text{\eqref{eq:ecrti}}}{=} \pro(\ti) \ecrti, \quad \forall \r \in \R, \quad \forall \ti \in \Ti, \quad \forall \i \in \I. \label{potential_prove}
\end{align}
Thus, $\potential (q)$ satisfies \eqref{potential_def} with $\gamma(\ti)=\pro(\ti)$ for any $\ti \in \Ti$ and any $\i \in \I$.  \QEDA

\emph{Proof of Lemma \ref{lemma:potential_convex}.} Since each $\cesfun(\we(\t))$ is differentiable in $\we(\t)$, we know that $\wpotential(\w)$ is twice differentiable with respect to $\w$. The first order partial derivative of $\wpotential(\w)$ with respect to $\we(\t)$ can be written as: $\frac{\partial \wpotential(\w)}{\partial \we(\t)}= \sum_{\s \in \S} \pi(\s, \t) \cesfun\(\we\(\t\)\)$ for any $\e \in \E$, and any $\t \in \T$. Also, the second order derivative of $\wpotential(\w)$ can be written as follows:
\begin{equation*}
\begin{split}
&\frac{\partial^2 \wpotential(\w)}{\partial \we(\t) \partial \w_{\e^{'}}(\t^{'})}=
\left\{
\begin{array}{ll}
\sum_{\s \in \S} \pi \(\s, \t \) \frac{d \cesfun\(\we\(\t\)\)}{d \we\(\t\)}, & \text{if $\e=\e^{'}$ and $\t=\t^{'}$,}\\
0, & \text{otherwise,}
\end{array}
\right. \quad \forall \e, \e^{'} \in \E, \quad \forall \t, \t^{'} \in \T.
\end{split}
\end{equation*}
Since for any $\e \in \E$ and any $\s \in \S$, $\cesfun(\we)$ is increasing in $\we$, $\sum_{\s \in \S} \pi \(\s, \t \) \frac{d \cesfun\(\we\(\t\)\)}{d \we\(\t\)}>0$. Thus, the Hessian matrix of $\wpotential(\w)$ has positive elements on the diagonal and 0 in all other entries, i.e. it is positive definite. Therefore, $\wpotential(\w)$ is strictly convex in $\w$.  \QEDA

\emph{Proof of Theorem \ref{q_opt}.} We first show that any minimum of \eqref{eq:potential_opt} is a BWE. The Lagrangian of \eqref{eq:potential_opt} is given by \eqref{lagrangian}, where $\mu=(\alti)_{\ti \in \Ti, \i \in \I}$ and $\nu=(\betir)_{\r \in \R, \ti \in \Ti, \i \in \I}$ are Lagrange multipliers associated with the constraints \eqref{sub:demand} and \eqref{sub:non-negative}, respectively. For any optimal solution $q$, there must exist $\mu$ and $\nu$ such that $\(\q, \mu, \nu\)$ satisfies the following Karush-Kuhn-Tucker (KKT) conditions:
\begin{taggedsubequations}{KKT}\label{KKT}
\begin{align}
\frac{\partial \lag}{\partial \qtir} &=  \frac{\partial \potential }{\partial \qtir}-\alti - \betir =0, \quad &\forall \r \in \R, \quad \forall \ti \in \Ti, \quad \forall \i \in \I, \label{first_order}\\
\betir \qtir &= 0, \quad &\forall \r \in \R, \quad \forall \ti \in \Ti, \quad \forall \i \in \I, \label{com_slack}\\
\betir &\geq 0, \quad &\forall \r \in \R, \quad \forall \ti \in \Ti, \quad \forall \i \in \I. \label{beta_positive}
\end{align}
\end{taggedsubequations}Using \eqref{potential_prove} and \eqref{first_order}, we have $\frac{\partial \potential(q) }{\partial \qtir} = \pro(\ti) \ecrti= \alti +\betir$ for any $\r \in \R$, and $\ti \in \Ti$, $\i \in \I$. From \eqref{com_slack}, we see that for any $\r \in \R$, and $\ti \in \Ti$, $\i \in \I$, if $\qtir>0$, the corresponding Lagrange multiplier $\betir=0$, and $\pro(\ti) \ecrti=\alti$. However, if $\qtir =0$, then $\pro(\ti) \ecrti=\alti+\betir \geq \alti$. Thus, for any $\r \in \R$, and $\ti \in \Ti$, $\i \in \I$:
\begin{equation*}
\begin{split}
\qtir>0 \quad \Rightarrow \quad \pro(\ti) \ecrti=\alti \leq \alti +\nu_{r^{'}}^{\ti} =\pro(\ti) \ecrpti, \quad \forall \r^{'} \in \R.
\end{split}
\end{equation*}
From \eqref{eq:BWE_fun}, we conclude that an optimal solution of \eqref{eq:potential_opt} is a BWE.

Next, we show that any BWE $\qwe$ of the game $\game$ is an optimal solution of \eqref{eq:potential_opt}. Consider a pair of Lagrange multipliers $\bar{\mu}$ (resp. $\bar{\nu})$) corresponding to the constraints (4a) (resp. (4b)), where $\bar{\mu}^{\ti }=\min_{\r \in \R}\pro(\ti) \ecrtiwe$ and $\bar{\nu}_\r^{\ti}=\pro(\ti) \ecrtiwe-\bar{\mu}^{\ti }$. We can easily check that \eqref{first_order} and \eqref{beta_positive} are satisfied by $(\qwe, \bar{\mu}, \bar{\nu})$. Since $\qwe$ is a BWE, we know from \eqref{eq:BWE_fun} that for a route $\r \in \R$, and $\ti \in \Ti$, $\i \in \I$, if $\qwetir>0$, then $\ecrtiwe=\min_{\r \in \R} \ecrtiwe$ and consequently $\bar{\nu}_{\r}^{\ti}=\pro(\ti) \ecrtiwe-\bar{\mu}^{\ti }=0.$ This implies that \eqref{com_slack} is also satisfied by $(\qwe, \bar{\mu}, \bar{\nu})$. Noting that $\potential(\q) \equiv \wpotential(\w)$, where the induced edge load $\w$ is linear in $\q$ (see \eqref{eq:q_w}), and that $\wpotential(\w)$ is strictly convex in $\w$ (Lemma \ref{lemma:potential_convex}), we conclude that $\potential(\q)$ is a convex function of $\q$. Furthermore, since $\Q(\lamb)$ is a convex polytope, \eqref{eq:potential_opt} is a convex problem. Thus, KKT conditions are also sufficient for optimality, and any BWE $\qwe$ is an optimal solution of \eqref{eq:potential_opt}. 

Finally, for any $\lamb$, we can use equations \eqref{eq:q_w} and \eqref{eq:potential_w} to re-express \eqref{eq:potential_opt} as an optimization problem whose solution gives an equilibrium edge load $\wwe(\lamb)$:
\begin{equation}\label{opt_w}
\begin{split}
\min_{\w, \q} &\quad \wpotential(\w)\\
s.t. &\quad \q \in \Q(\lamb), \quad \we(\t)=\sum_{\r \ni \e} \sum_{\i \in \I}\qtir, \quad \forall \t \in \T, \quad \forall \e \in \E.
\end{split}
\end{equation}
Clearly, the feasible set of the above problem is a convex polytope. From Lemma \ref{lemma:potential_convex}, $\wpotential(\w)$ is strictly convex in $\w$. Therefore, the equilibrium edge load $\wwe(\lamb)$ is unique.  \QEDA

\begin{lem}{(Theorem 2 in \cite{wachsmuth2013licq})}\label{licq_lemma}
The Lagrange multiplies $\mu^{*}$ and $\nu^{*}$ associated with any $\qwe \in \Qwe(\lamb)$ at the optimum of \eqref{eq:potential_opt} are unique if and only if the LICQ condition is satisfied in that the gradients of the set of tight constraints in \eqref{sub:demand}-\eqref{sub:non-negative} at the optimum are linearly independent.
\end{lem}

\emph{Proof of Lemma \ref{unique_lag}.}
Let the set of constraints that are tight at optimum of \eqref{eq:potential_opt} in \eqref{sub:non-negative} be denoted as $\mathcal{B}$. Assume for the sake of contradiction that LICQ does not hold, i.e. the set of equality constraints \eqref{sub:demand} and the elements in the set $\mathcal{B}$ are linearly dependent. Now, note that the constraint sets \eqref{sub:demand} and \eqref{sub:non-negative} are each comprised of linearly independent affine functions. Hence, there must exist a type $\bar{\t}^\i$ such that the gradient of the corresponding equality constraint (i.e. $\sum_{\r \in \R} \q_\r^{\i*}(\bar{\t}^\i) =\lambi \D$) is linearly dependent with the elements in the set $\mathcal{B}$, which implies that $\q_\r^{\i*}(\bar{\t}^\i)=0, \forall \r \in \R$. However, this violates the equality constraint in \eqref{sub:demand} as $\sum_{\r \in \R} \q_\r^{\i*}(\bar{\t}^\i)=\lambi \D \neq 0$, and we arrive at a contradiction.

Since LICQ holds, for any equilibrium strategy profile $\qwe \in \Qwe(\lamb)$, the corresponding $\mu^{*}$ and $\nu^{*}$ must be unique. Following the proof of Theorem \ref{q_opt}, we conclude that for any $\qwe \in \Qwe(\lamb)$, $\(\qwe, \mu^{*}, \nu^{*}\)$ satisfies the KKT conditions, and $\mu^{\ti*}$ and $\nu^{\ti*}_\r$ can be written as \eqref{define_al} and \eqref{define_beta}, respectively. 

Finally, noting that the equilibrium edge load is unique (Theorem \ref{q_opt}), $\mu^{*}$ and $\nu^{*}$ in \eqref{define_al}-\eqref{define_beta} are thus unique in equilibrium.  \QEDA


\emph{Proof of Proposition \ref{Lprime}.}
\underline{Step I}: We show that any $\q \in \Q(\lamb)$ induces a route flow $\l$ that satisfies \eqref{sub:balance}-\eqref{sub:popu_i}. From \eqref{eq:load}, we obtain that for any $\ti, \titil \in \Ti$, any $\tmi, \tmitil \in \Tmi$, and any $\i \in I$, $\l$ satisfies \eqref{sub:balance}:
\begin{equation*}
\begin{split}
&\lr(\ti, \tmi)-\lr(\titil, \tmi)= \qtir+\sum_{\j \in \I \setminus \{\i\}} \qtjr-\qtirtil-\sum_{\j \in \I \setminus \{\i\}} \qtjr\\
=&\qtir+\sum_{\j \in \I \setminus \{\i\}} \qtjrtil-\qtirtil-\sum_{\j \in \I \setminus \{\i\}} \qtjrtil
= \lr(\ti, \tmitil)-\lr(\titil, \tmitil).
\end{split}
\end{equation*}
From \eqref{sub:demand} and \eqref{sub:non-negative}, we can directly conclude that $\l$ must also satisfy \eqref{sub:ldemand} and \eqref{sub:l_positive}. Additionally, 
\begin{align*}
&\D-\sum_{\r \in \R}\min_{\ti \in \Ti}\lr(\ti, \tmi) \stackrel{\eqref{eq:load}}{=}\D -\sum_{\r \in \R} \sum_{\j \in \I \setminus \{\i\}}\qtjr-\sum_{\r \in \R}\min_
{\ti \in \Ti} \qtir\\ \stackrel{\eqref{sub:demand}}{=}& \D-\sum_{\j \in \I \setminus \{\i\}}\lamb^j \D -\sum_{\r \in \R}\min_
{\ti \in \Ti} \qtir=\lambi \D- \sum_{\r \in \R}\min_
{\ti \in \Ti} \qtir \stackrel{\eqref{sub:non-negative}}{\leq} \lambi \D,  \text{ } \quad \forall \tmi \in \Tmi, \text{ } \forall \i \in \I. 
\end{align*}
Therefore, $\l$ satisfies \eqref{sub:popu_i}. Thus, any feasible route flow must satisfy \eqref{sub:balance}-\eqref{sub:popu_i}.

\underline{Step II}: Next, we show that for any route flow $\l \in \F(\lamb)$ (i.e. $\l$ that satisfies constraints \eqref{sub:balance}-\eqref{sub:popu_i}), the set of feasible strategies that induce $\l$ can be characterized by \eqref{eq_rep}. For any route $\r \in \R$, the linear system of equations \eqref{eq:load} has $\prod_{\i \in \I} |\Ti|$ equations in $\sum_{\i \in \I} |\Ti|$ variables. Note that for any given $\that =\(\tihat\)_{\i \in \I} \in \T$, the following equations are linearly independent:
\begin{equation}\label{eq:indep}
\begin{split}
\sum_{\i \in \I} \q^\i_\r(\tihat) &=\lr(\that),\\
\qtir+\sum_{\j \in \I \setminus \{\i\}} \q_\r^j(\widehat{t}^j)&=\lr(\ti, \tmihat), \quad \forall \ti \in \Ti \setminus \{\tihat\}, \quad \forall \i \in \I.
\end{split}
\end{equation}
We then show that given any $\t \in \T$, $\sum_{\i \in \I} \qtir=\lr(\t)$ is a linear combination of the equations in \eqref{eq:indep}. Following \eqref{sub:balance}, we can write:
\begin{align*}
&\sum_{\i \in \I} \lr(\ti, \tmihat)-(|\I|-1)\lr(\that) =\lr(\tone, \widehat{\t}^{-1})+\lr(\ttwo, \widehat{\t}^{-2}) +\sum_{\i=3}^{|\I|} \lr(\ti, \tmihat)-(|\I|-1)\lr(\that)\\ \stackrel{\eqref{sub:balance}}{=}&\lr(\tone, \ttwo, \widehat{\t}^{-1-2})+\lr(\that)+\sum_{\i=3}^{|\I|} \lr(\ti, \tmihat)-(|\I|-1) \lr(\that)\\
=&\lr(\tone, \ttwo, \widehat{\t}^{-1-2})+\sum_{\i=3}^{|\I|} \lr(\ti, \tmihat)- (|\I|-2) \lr(\that),
\end{align*}
where $\widehat{\t}^{-1-2}=(\widehat{\t}^3, \cdots, \widehat{\t}^{|\I|})$. We apply the same procedure iteratively for another $|\I|-2$ times:
\begin{align}\label{iterative}
\sum_{\i \in \I} \lr(\ti, \tmihat)-(|\I|-1)\lr(\that)
&=\lr(\t), \quad \forall \t \in \T.
\end{align}
Now for any $\r \in \R$ and $\t \in \T$, we can write $\sum_{\i \in \I} \qtir=\sum_{\i \in \I} \(\qtir+\sum_{\j \in \I \setminus \{\i\}} \q_\r^j(\widehat{t}^j)\)- (|\I|-1) \sum_{\i \in \I} \q^\i_\r(\tihat)\stackrel{\text{\eqref{eq:indep}}}{=}\sum_{\i \in \I} \lr(\ti, \tmihat)-(|\I|-1)\lr(\that) \stackrel{\eqref{iterative}}{=}\lr(\t).$
Thus, for any $\r \in \R$, the linear system \eqref{eq:load} is comprised of $\sum_{\i \in \I} |\Ti|$ variables, and any constraint can indeed be expressed as a linear combination of $\sum_{\i \in \I} |\Ti| -|\I|+1$ independent equations in \eqref{eq:indep}. From the rank-nullity theorem, we conclude that the dimension of null space of this linear map is $|\I|-1$. Then, for any $\r \in \R$, any $\i \in \I$, setting $\q^\i_\r(\tihat)=\chi_\r^\i$, any solution of \eqref{eq:load} can be expressed as \eqref{eq_rep}, where $\that \in \T$ is an arbitrary type profile. Additionally, $\sum_{\i \in \I}\chi_\r^\i=\sum_{\i \in \I}\q^\i_\r(\tihat)=\lr(\that)$. Thus, $\chi$ satisfies \eqref{sub:sum_demand}, i.e. for each $\r \in \R$, out of the $|\I|$ variables in $\{\chi_\r^\i\}_{\i \in \I}$, $|\I|-1$ are free, and the remaining one is obtained from \eqref{sub:sum_demand}. We can conclude that the strategy profile $\q$ as defined in \eqref{eq_rep} induces the route flow $\l$. It remains to be shown that if $\q$ is a feasible strategy profile, $\chi$ must satisfy \eqref{sub:x_sum} and \eqref{sub:x_bound} as well. Since $\q$ satisfies \eqref{sub:demand}, we obtain that $\lambi \D \stackrel{\eqref{sub:demand}}{=} \sum_{\r \in \R} \q^{\i}_\r(\ti)\stackrel{\eqref{eq_rep}}{=} \sum_{\r \in \R} \(\lr(\ti, \tmihat)-\lr(\tihat, \tmihat)+\chi_\r^\i\)\stackrel{\text{\eqref{sub:ldemand}}}{=} \sum_{\r \in \R} \chi_\r^\i$ for any $\i \in \I$, i.e. $\chi$ satisfies \eqref{sub:x_sum}:
Additionally, from \eqref{sub:non-negative}, $0 \leq q^{\i}_\r(\ti)\stackrel{\eqref{eq_rep}}{=}\lr(\ti, \tmihat)-\lr(\tihat, \tmihat)+\chi_\r^{\i}$ for any $\r \in \R$ and any $\ti \in \Ti$. Thus, $\chi_\r^{\i} \geq \max_{\ti \in \Ti} \left\{\lr(\tihat, \tmihat)-\lr(\ti, \tmihat)\right\}$, i.e. $\chi$ satisfies \eqref{sub:x_bound}.

Step III: Finally, we show that the set of $\chi$ satisfying \eqref{X} is non-empty, i.e., any $\l \in \F(\lamb)$ can be induced by at least one feasible strategy profile $\q$. Consider any $\l \in \F(\lamb)$, we explicitly construct the following $\chi$, and show that such $\chi$ satisfies \eqref{X}:
\begin{align}\label{x_example}
\noindent\chi_\r^\i=\gamma_\r \cdot \(\lambi \D -\sum_{\r \in \R} \max_{\ti \in \Ti} \(\lr(\that)-\lr(\ti, \tmihat)\)\) + \max_{\ti \in \Ti} \(\lr(\that)-\lr(\ti, \tmihat)\), \text{   } \forall \r \in \R, \text{  } \forall \i \in \I, 
\end{align}
where $\that$ is any arbitrary type profile, and $\gamma_\r=\frac{\lr(\that)-\sum_{\i \in \I} \max_{\ti \in \Ti} \(\lr(\that)-\lr(\ti, \tmihat)\)}{\sum_{\r \in \R} \left[\lr(\that)-\sum_{\i \in \I} \max_{\ti \in \Ti} \(\lr(\that)-\lr(\ti, \tmihat)\)\right]}$ if $\sum_{\r \in \R} \left[\lr(\that)-\sum_{\i \in \I} \max_{\ti \in \Ti} \(\lr(\that)-\lr(\ti, \tmihat)\)\right]\neq 0$ and 0 otherwise.

First, we check that the $\(\chi_\r^\i\)_{\r \in \R, \i \in \I}$ as defined in \eqref{x_example} satisfies \eqref{sub:x_bound}. Note that $\gamma_\r\geq 0$. To see this, since for any $\t \in \T$, $\sum_{\i \in \I} \lr(\ti, \tmihat)-(|\I|-1)\lr(\that)\stackrel{\eqref{iterative}}{=}\lr(\t) \geq 0$, we know that $\min_{\t \in \T}\sum_{\i \in \I}  \lr(\ti, \tmihat)-(|\I|-1)\lr(\that) =\min_{\t \in \T} \lr(\t)\geq 0$. Thus, for any $\r \in \R$, we obtain:
\begin{align}
\lr(\that)-\sum_{\i \in \I} \max_{\ti \in \Ti} \(\lr(\that)-\lr(\ti, \tmihat)\) 
&=\min_{\t \in \T} \sum_{\i \in \I}  \lr(\ti, \tmihat)- (|\I|-1)\lr(\that)=\min_{\t \in \T} \lr(\t)\geq 0. \label{min_t}
\end{align}
Hence, we can conclude that $\gamma_\r \geq0$. Next, $\lambi \D -\sum_{\r \in \R} \max_{\ti \in \Ti} \(\lr(\that)-\lr(\ti, \tmihat)\) \stackrel{\eqref{sub:ldemand}}{=}\lambi \D -\(\D-\sum_{\r \in \R} \min_{\ti \in \Ti}\lr(\ti, \tmihat)\) \stackrel{\eqref{sub:popu_i}}{\geq} 0$. Using the above inequalities, we obtain that $\chi_\r^\i$ as considered in \eqref{x_example} satisfies \eqref{sub:x_bound}. 

Second, we check $\chi_\r^\i$ satisfies \eqref{sub:x_sum}. If $\sum_{\r \in \R} \left[\lr(\that)-\sum_{\i \in \I} \max_{\ti \in \Ti} \(\lr(\that)-\lr(\ti, \tmihat)\)\right] >0$, then:
\begin{align*}
\sum_{\r \in \R} \chi_\r^\i &=\sum_{\r \in \R} \gamma_\r \cdot \(\lambi \D -\sum_{\r \in \R} \max_{\ti \in \Ti} \(\lr(\that)-\lr(\ti, \tmihat)\)\) +\sum_{\r \in \R} \max_{\ti \in \Ti} \(\lr(\that)-\lr(\ti, \tmihat)\)\\
&=\(\lambi \D -\sum_{\r \in \R} \max_{\ti \in \Ti} \(\lr(\that)-\lr(\ti, \tmihat)\)\)+\sum_{\r \in \R} \max_{\ti \in \Ti} \(\lr(\that)-\lr(\ti, \tmihat)\)=\lambi \D. 
\end{align*}
On the other hand, if $\sum_{\r \in \R} \left[\lr(\that)-\sum_{\i \in \I} \max_{\ti \in \Ti} \(\lr(\that)-\lr(\ti, \tmihat)\)\right]=0$, we obtain that:
\begin{footnotesize}
\begin{align*}
0&=\sum_{\r \in \R} \left[\lr(\that)-\sum_{\i \in \I} \max_{\ti \in \Ti} \(\lr(\that)-\lr(\ti, \tmihat)\)\right]\stackrel{\eqref{sub:ldemand}}{=} \D- \sum_{\i \in \I} \(\sum_{\r \in \R}\max_{\ti \in \Ti} \(\lr(\that)-\lr(\ti, \tmihat)\)\) \\
&\stackrel{\eqref{sub:popu_i}}{\geq} \D-\sum_{\i \in \I} \lambi \D =0,
\end{align*}
\end{footnotesize}which implies that for any $\i \in \I$, $\sum_{\r \in \R}\max_{\ti \in \Ti} \(\lr(\that)-\lr(\ti, \tmihat)\)=\lambi \D$. Since in this case, $\gamma_\r=0$, we can conclude that $\sum_{\r \in \R}\chi_\r^\i= \sum_{\r \in \R}\max_{\ti \in \Ti} \(\lr(\that)-\lr(\ti, \tmihat)\)=\lambi \D$, i.e. $\chi$ satisfies \eqref{sub:x_sum}. 

Finally, $\chi_\r^\i$ also satisfies \eqref{sub:sum_demand}. If $\sum_{\r \in \R} \left[\lr(\that)-\sum_{\i \in \I} \max_{\ti \in \Ti} \(\lr(\that)-\lr(\ti, \tmihat)\)\right] >0$, we have:
\begin{align*}
\sum_{\i \in \I} \chi_\r^\i &= \gamma_\r \cdot \sum_{\i \in \I} \(\lambi \D -\sum_{\r \in \R} \max_{\ti \in \Ti} \(\lr(\that)-\lr(\ti, \tmihat)\)\) +\sum_{\i \in \I} \max_{\ti \in \Ti} \(\lr(\that)-\lr(\ti, \tmihat)\)\\
&= \gamma_\r \cdot \(\D -\sum_{\i \in \I}\sum_{\r \in \R} \max_{\ti \in \Ti} \(\lr(\that)-\lr(\ti, \tmihat)\)\) +\sum_{\i \in \I} \max_{\ti \in \Ti} \(\lr(\that)-\lr(\ti, \tmihat)\)\\
&\stackrel{\eqref{sub:ldemand}}{=}\gamma_\r \cdot \(\sum_{\r \in \R} \left[\lr(\that)-\sum_{\i \in \I} \max_{\ti \in \Ti} \(\lr(\that)-\lr(\ti, \tmihat)\)\right]\)+\sum_{\i \in \I} \max_{\ti \in \Ti} \(\lr(\that)-\lr(\ti, \tmihat)\)\\
&=\lr(\that)-\sum_{\i \in \I} \max_{\ti \in \Ti} \(\lr(\that)-\lr(\ti, \tmihat)\)+\sum_{\i \in \I} \max_{\ti \in \Ti} \(\lr(\that)-\lr(\ti, \tmihat)\)=\lr(\that).
\end{align*}
If $\sum_{\r \in \R} \left[\lr(\that)-\sum_{\i \in \I} \max_{\ti \in \Ti} \(\lr(\that)-\lr(\ti, \tmihat)\)\right] =0$, then we have 
\[0=\sum_{\r \in \R} \left[\lr(\that)-\sum_{\i \in \I} \max_{\ti \in \Ti} \(\lr(\that)-\lr(\ti, \tmihat)\)\right] \stackrel{\eqref{min_t}}{=}\sum_{\r \in \R} \min_{\t \in \T} \lr(\t) \geq 0,\] which implies that for any $\r \in \R$, $\min_{\t \in \T}\lr(\t)=0$. In this case, $\gamma_r=0$, and thus $\sum_{\i \in \I} \chi_\r^\i=\sum_{\i \in \I} \max_{\ti \in \Ti} \(\lr(\that)-\lr(\ti, \tmihat)\)\stackrel{\eqref{min_t}}=\lr(\that)-\min_{\t \in \T} \lr(\t)=\lr(\that)$, i.e. $\chi$ satisfies \eqref{sub:sum_demand}.

Therefore, if $\l$ satisfies \eqref{sub:balance}-\eqref{sub:popu_i}, we can conclude that the $\chi$ in \eqref{x_example} satisfies \eqref{X}, i.e. the set of $\chi$ satisfying \eqref{X} is non-empty. We already showed in Step II that $\q$ as defined in \eqref{eq_rep} with parameter $\chi$ satisfying \eqref{X} is a feasible strategy profile, and $\q$ induces $\l$. 
Therefore, if $\l$ satisfies  \eqref{sub:balance}-\eqref{sub:popu_i}, there exists a feasible $\q$ that induces $\l$, i.e. any $\l\in \F(\lamb)$ is a feasible route flow. 

In summary, we have shown that any feasible route flow satisfies \eqref{eq:Lprime} (Step I); For any $\l$ that satisfies \eqref{eq:Lprime}, the set of feasible strategy profiles that induce $\l$ can be written in \eqref{eq_rep}-\eqref{X} (Step II); Such set is non-empty, and hence $\l$ is feasible (Step III). We can thus conclude that the set of feasible route flows is $\F(\lamb)$, and the set of feasible strategies that induce $\l$ is as in \eqref{eq_rep}-\eqref{X}. \QEDA

\emph{Proof of Proposition \ref{l_opt}.}
From Proposition \ref{Lprime}, we know that the set of feasible route flows is the set $\F(\lamb)$ characterized by \eqref{sub:balance}-\eqref{sub:popu_i}. Additionally, the weighted potential function in \eqref{eq:potential_q} can be equivalently written as a function of $\l$ given by \eqref{eq:potential_l}. Therefore, the minimum of \eqref{opt_l} is equal to that in \eqref{eq:potential_opt}, and the set of optimal solutions is the set of equilibrium route flows.   \QEDA

\section{Supplementary Material for Section \ref{equilibrium_regime}}\label{appendix_B}
\begin{lem}\label{unique_wdagij}
The route flows $\ldagij \in \Ldagij$ induce a unique edge load $\wdagij$. 
\end{lem}
\emph{Proof of Lemma \ref{unique_wdagij}}
Following \eqref{eq:q_w} and \eqref{opt_l}, any edge load $\wdagij$ induced by route flows in $\Ldagij$ (which we defined as optimal solution set of \eqref{eq:basic_opt_ij}) is an optimal solution of the following optimization problem:
\begin{align*}
\min_{\w}\quad & \wpotential(\w), \\
s.t. \quad & \we(\t)=\sum_{\r \ni \e} \lr(\t), \quad \forall \t \in \T, \quad \forall \e \in \E,\\
& \text{$\l$ satisfies \eqref{sub:balance}, \eqref{sub:ldemand}, \eqref{sub:l_positive}, \eqref{prime:popu_i}$\setminus\{\i, \j\}$, \eqref{extra}.}
\end{align*}
The constraints \eqref{sub:balance}, \eqref{sub:ldemand}, \eqref{sub:l_positive} are linear constraints. Following from \eqref{affine_popu_i}, constraints \eqref{prime:popu_i}$\setminus\{\i, \j\}$, \eqref{extra} are each equivalent to a set of linear constraints. Additionally, $\w$ is a linear function of $\l$, thus the feasible set of $\w$ in this optimization problem must also be a convex polytope. Since $\wpotential(\w)$ is a strictly convex function in $\w$, the optimal solution $\wdagij$ is unique.  \QEDA

\emph{Proof of Lemma \ref{lemma:lambl_lambup}.}
First, we show that both thresholds $\lambli$ and $\lambupi$ belong to the interval $[0,1-|\lambmimj|]$. Since $\lambli$ is attainable on the set $\Ldagij$, there exists $\ldagijtil \in \Ldagij$ such that:
\begin{equation*}
\begin{split}
\lambli = \frac{1}{\D}\widehat{\md}^{\i}(\ldagijtil) \stackrel{\text{\eqref{widehatj}}}{=}\frac{1}{\D}\(\D-\sum_{\r \in \R} \min_{\ti \in \Ti} \ldagijtil_\r\(\ti, \tmi\)\) \geq\frac{1}{\D}\(\D-\sum_{\r \in \R} \ldagijtil_\r\(\tihat, \tmi\)\)\stackrel{\eqref{sub:ldemand}}{=}0. 
\end{split}
\end{equation*}
Similarly, we can check that $\lambupi \leq 1-|\lambmimj|$. 

Additionally, we know for any $\ldagij \in \Ldagij$:
\begin{align*}
\lambupi \stackrel{\eqref{pairwise_threshold}}{\geq} \frac{1}{\D} \left\{\(1-|\lambmimj|\)\D-\widehat{\md}^{\j}(\ldagij)\right\} \stackrel{\eqref{extra}}{\geq} \frac{1}{\D}\mdl^{\i}(\ldagij) \stackrel{\eqref{pairwise_threshold}}{\geq} \lambli.
\end{align*}

Therefore, $0 \leq \lambli \leq \lambupi \leq 1-|\lambmimj|$.  \QEDA

For any two populations $\i, \j \in \I$, we can compute the threshold $\lambli$ as follows:
\begin{equation}\label{linear_program_lambli}
\begin{split}
\min \quad &y \\
s.t. \quad & \D- \sum_{\r \in \R} \lr(\ti_\r, \tmihat) \leq y \cdot \D, \quad \forall \ti_1 \in \Ti, \dots, \forall \ti_{|\R|} \in \Ti,\\
& \ldagij \in \Ldagij,
\end{split}
\end{equation}
where $\Ldagij$ is the polytope defined in \eqref{drop_i_j}. Therefore, \eqref{linear_program_lambli} is a linear programming.  Analogously, the threshold $\lambupi$ is the optimal value of the following linear program:
\begin{equation}\label{linear_program_lambupi}
\begin{split}
\max \quad &y \\
s.t. \quad & -|\lambmimj|\D+ \sum_{\r \in \R} \lr(\tj_\r, \tmjhat) \geq y \cdot \D, \quad \forall \tj_1 \in \Tj,\dots \forall \tj_{|\R|} \in \Tj,\\
& \ldagij \in \Ldagij.
\end{split}
\end{equation}

\emph{Proof of Theorem \ref{l_behavior}.}
$\left[\text{Regime } \Reoneij \right]$: First, we show by contradiction that the constraint (\ref{prime:popu_i}$_i$) is tight for any equilibrium route flow. Assume that for a given $\lamb \in \Reoneij$, there exists an equilibrium route flow $\lwe$ such that (\ref{prime:popu_i}$_\i$) is not tight. From Proposition \ref{l_opt}, we know that $\lwe$ is an optimal solution of \eqref{opt_l}. Since $\eqref{opt_l}$ is a convex optimization problem, $\lwe$ is still a minimizer of $\lpotential(\l)$ if we drop the constraint (\ref{prime:popu_i}$_i$). Additionally, the constraints (\ref{prime:popu_i}$_i$) and (\ref{prime:popu_i}$_{j}$) implies that $\lwe$ must also satisfy \eqref{extra}. Thus, such $\lwe$ is an optimal solution of the following problem:
\begin{equation}\label{eq_l_extra}
\min_{\l} \quad  \lpotential(\l), \quad s.t. \quad \text{\eqref{sub:balance}, \eqref{sub:ldemand}, \eqref{sub:l_positive}, \eqref{extra}, and \eqref{prime:popu_i}$\setminus \{\i\}$}. 
\end{equation}
Moreover, the threshold $\lambupi$ defined in \eqref{pairwise_threshold} is attained by a route flow, say $\ldagijtil$, in the set $\Ldagij$. Thus, we can write:
$1-|\lambmimj|-\frac{1}{\D}\widehat{\md}^{\j}(\ldagijtil) = \lambupi \stackrel{\text{(Lemma \ref{lemma:lambl_lambup})}}{\geq} \lambli \stackrel{\eqref{def_reg_one}}{>}\lambi.$ 
Rearranging, we obtain: $\frac{1}{\D}\widehat{\md}^{\j}(\ldagijtil)  < 1-|\lambmimj| -\lambi=\lambj$, and so such $\ldagijtil$ also satisfies (\ref{prime:popu_i}$_{j}$). Since $\ldagijtil$ is an optimal solution of \eqref{drop_i_j}, which minimizes the same objective function as \eqref{eq_l_extra} but without the constraint (\ref{prime:popu_i}$_{j}$), we thus know that $\ldagijtil$ is also an optimal solution in \eqref{eq_l_extra}. Since the induced edge load is unique, we can conclude that the edge load induced by $\lwe$ must be identical to that induced by $\ldagijtil$, which is $\wdagij$. Then, from \eqref{drop_i_j}, we have $\lwe \in \Ldagij$. Therefore, from \eqref{pairwise_threshold}, we can write $\lambli \leq \frac{1}{\D} \widehat{\md}^{\i}(\lwe)$. Since we assumed that (\ref{prime:popu_i}$_{i}$) is not binding in equilibrium, we obtain: $\frac{1}{\D} \mdl^{\i}(\lwe)< \lambi < \lambli\leq \frac{1}{\D} \mdl^\i(\lwe)$, which is a contradiction. Thus, (\ref{prime:popu_i}$_{i}$) must be tight in equilibrium for any $\lamb$ in regime $\Reoneij$.

Finally, following the tightness of (\ref{prime:popu_i}$_{i}$) at optimum of \eqref{opt_l}, by rearranging the constraint \eqref{extra} in \eqref{eq:l_behavior}, we have $\mdl^{\j}(\lwe) \leq \(1-|\lambmimj|\) \D- \mdl^{\i}(\lwe)  =\lamb^\j \D$. 
Thus, (\ref{prime:popu_i}$_{j}$) is guaranteed to hold in Regime $\Reoneij$ given the constraint \eqref{extra} and the fact that (\ref{prime:popu_i}$_{i}$) is tight at the optimum of \eqref{opt_l}. Hence, (\ref{prime:popu_i}$_{j}$) can be dropped in \eqref{opt_l} without changing the optimal solution set. 

$\left[\text{Regime } \Rethreeij \right]$: Analogous to the proof given for regime $\Reoneij$, we can argue that constraint (\ref{prime:popu_i}$_{j}$) is tight in any equilibrium for any $\lamb$ in regime $\Rethreeij$. By imposing constraint \eqref{extra}, (\ref{prime:popu_i}$_i$) can be dropped from the constraint set in \eqref{opt_l} without changing the optimal solution set.

$\left[\text{Regime } \Retwoij \right]$:
To study this regime, we need two additional thresholds 
\[\lamblip \deleq \frac{1}{\D} \max_{\ldagij \in \Ldagij}\left\{\widehat{\md}^{\i}(\ldagij)\right\}, \text{ and } \lambupip \deleq \frac{1}{\D} \min_{\ldagij
 \in \Ldagij}\left\{\(1-|\lambmimj|\)\D-\widehat{\md}^{\j}(\ldagij)\right\}.\]
 From \eqref{pairwise_threshold}, we can check that $\lambli \leq \lamblip$, and $\lambupip \leq \lambupi$.

For any $\lambi \in [\lambli, \lamblip]$, we argue that $\Lwelamb \subseteq \Ldagij$. Since the set $\Ldagij$ as defined by \eqref{drop_i_j} is a bounded polytope, and $\lambli$ (resp. $\lamblip$) is the minimum (resp. maximum) value of the continuous function $\mdl^{\i}(\ldagij)$ on $\Ldagij$, we know from the mean value theorem that there exists a $\ldagijtil \in \Ldagij$ satisfying: $\lamb^\i =\frac{1}{\D} \mdl^{\i}(\ldagijtil)$. Such $\ldagijtil$ also satisfies constraint (\ref{prime:popu_i}$_{j}$). Therefore, $\ldagijtil$ satisfies all the constraints in \eqref{eq:Lprime}, and minimizes $\lpotential(\l)$. So $\ldagijtil$ is an equilibrium route flow, which implies that $\Lwelamb \cap \Ldagij \neq \emptyset$. Since the equilibrium edge load vector is unique, and the edge load induced by $\ldagijtil$ is $\wdagij$, we must have $\wwelamb=\wdagij$. Furthermore, from \eqref{drop_i_j}, $\Ldagij$ is a superset of all feasible route flows that can induce $\wdagij$. Therefore, $\Lwelamb \subseteq \Ldagij$ for any $\lamb^\i \in [\lambli, \lamblip]$. Similarly, we can argue that for any $\lambi \in [\lambupip, \lambupi]$, $\Lwelamb \subseteq \Ldagij$.

To prove that $\Lwe(\lamb) \subseteq \Ldagij$ for any $\lamb$ in regime $\Retwoij$, we need to argue two cases $\lamblip \geq  \lambupip$ and $\lamblip < \lambupip$ separately. If $\lamblip \geq  \lambupip$, then $[\lambli, \lambupi] \subseteq [\lambli, \lamblip] \cup [\lambupip, \lambupi]$. Therefore, $\Lwelamb \subseteq \Ldagij$ for any $\lamb$ in regime $\Retwoij$. If $\lamblip< \lambupip$, for any $\lambi \in (\lamblip, \lambupip)$, we can check that any $\ldagij \in \Ldagij$ satisfies the constraint (\ref{prime:popu_i}$_i$): $\frac{1}{\D}\mdl^{\i}(\ldagij) \leq \lamblip<\lambi$. Additionally, since $\lambi < \lambupip \leq  1-|\lambmimj|-\frac{1}{\D} \mdl^{\j}(\ldagij)$, we know that $\frac{1}{\D} \mdl^{\j}(\ldagij) <1-|\lambmimj|-\lambi =\lambj$, i.e. $\ldagij$ also satisfies the constraint (\ref{prime:popu_i}$_{j}$). Thus, any $\ldagij \in \Ldagij$ is an equilibrium route flow, i.e. $\Lwelamb =\Ldagij$ for any $\lambi \in (\lamblip, \lambupip)$. Combined with the fact that $\Lwelamb \subseteq \Ldagij$ for any $\lamb \in [\lambli, \lamblip] \cup [\lambupip, \lambupi]$, we know that $\Lwelamb \subseteq \Ldagij$ for any $\lamb$ in regime $\Retwoij$. \QEDA

\begin{coro}\label{cor:strategy}
If the game $\game$ has a parallel-route network, then the equilibrium route flow $\lwe$ is unique. Moreover, if there are two populations, then the equilibrium strategy profile is unique in regime $\Reone^{12}$ or $\Rethree^{12}$, and can be written as follows:
\begin{subequations}
\begin{align}
\text{In regime $\Reoneij$: }\quad  \q^{1*}_\r(\tone)&=\lwer(\tone, \ttwohat)-\min_{\tonehat \in \Tone} \lwer(\tonehat, \ttwohat), \quad \forall \r \in \R, \quad \forall \tone \in \Tone, \label{q11}\\
\q^{2*}_\r(\ttwo)&=\min_{\tonehat \in \Tone} \lwer(\tonehat, \ttwo), \quad \forall \r \in \R, \quad \forall \ttwo \in \Ttwo, \label{q21}\\
\text{In regime $\Rethreeij$: }\quad  \q^{1*}_\r(\tone)&=\min_{\ttwohat \in \Ttwo} \lwer(\tone, \ttwohat), \quad \forall \r \in \R, \quad \forall \tone \in \Tone,\label{q_13}\\
\q^{2*}_\r(\ttwo)&=\lwer(\tonehat, \ttwo)-\min_{\ttwohat \in \Ttwo} \lwer(\tonehat, \ttwohat), \quad \forall \r \in \R, \quad \forall \ttwo \in \Ttwo,\label{q_23}
\end{align} 
\end{subequations}
where $\(\tonehat, \ttwohat\)$ is any type profile.
\end{coro}
\emph{Proof of Corollary \ref{cor:strategy}.}
Given a parallel route network, we immediately obtain the uniqueness of $\lwe$ from Theorem \ref{q_opt}. Then from Proposition \ref{Lprime}, any strategy profile that can induce $\lwe$ can be expressed as in \eqref{eq_rep}. In regime $\Reone^{12}$, we know from Theorem \ref{l_behavior} that the constraint (\ref{prime:popu_i}$_1$) is tight in equilibrium. Therefore, from \eqref{sub:x_sum} and \eqref{sub:x_bound}, we obtain:
\begin{align*}
\lamb^1 \D\stackrel{\eqref{sub:x_sum}}{=}  \sum_{\r \in \R} \chi_\r^1 \stackrel{\eqref{sub:x_bound}}{\geq} \sum_{\r \in \R}\max_{\tone\in \Tone} \(\lrwe(\tonehat, \ttwohat)-\lrwe(\tone, \ttwohat)\)\stackrel{\eqref{widehatj}}{=} \mdl^{1} (\lwe) =\lamb^1 \D.
\end{align*}
Thus, \eqref{sub:x_bound} is tight for any $\r \in \R$, i.e. $\chi_\r^1=\max_{\tone \in \Tone} \(\lrwe(\tonehat, \ttwohat)-\lrwe(\tone, \ttwohat)\)$. Additionally, from \eqref{sub:x_sum}, $\chi_\r^2=\min_{\tone \in \Tone} \lrwe(\tone, \ttwohat)$. Thus, $\chi$ as defined in \eqref{X} is unique. Following \eqref{eq_rep}, we can obtain the unique $\qwe$ as defined in \eqref{q11}-\eqref{q21}. Analogously, we can argue that the equilibrium strategy profile is also unique in regime $\Rethree^{12}$, and is written as in \eqref{q_13}-\eqref{q_23}. \QEDA

\emph{Proof of Proposition \ref{bathtub}.}
$\left[\text{Regime $\Reoneij$}\right]$: Consider any population size vector $\lamb \in \Reone^{ij}$, there exists a sufficiently small $\epsilon >0$ such that $\lamb^{'} =\lamb+\epsilon \zij \in \Reone^{ij}$, i.e. $\lamb^{i'}=\lambi+\epsilon>\lambi$,  $\lamb^{j'}=\lamb^j-\epsilon<\lamb^j$, and the sizes of all other populations remain unchanged. Consider any equilibrium route flow $\lwe(\lamb) \in \Lwe(\lamb)$ and any $\lwe(\lamb^{'}) \in \Lwe(\lamb^{'})$. We know from Theorem \ref{l_behavior} that constraint (\ref{prime:popu_i}$_i$) is tight in equilibrium, and thus $\lwe(\lamb)$ and $\lwe(\lamb^{'})$ satisfy: $\frac{1}{\D}\mdl^{\i}(\lwe(\lamb)) = \lambi< \lamb^{\i '} =\frac{1}{\D}\mdl^{\i}(\lwe(\lamb^{'}))$. Consequently, any equilibrium route flow $\lwe(\lamb)$ for size vector $\lamb$ is in the feasible domain of \eqref{eq:l_behavior} for size vector $\lamb^{'}$, but $\lwe(\lamb) \notin \Lwe(\lamb^{'})$, because $\mdl^{\i}\(\lwe(\lamb)\)=\lambi < \lamb^{\i'}$, i.e. the constraint (\ref{prime:popu_i}$_i$) is satisfied, but not tight. Since $\lwe(\lamb^{'}) \in \Lwe(\lamb^{'})$, we must have $\Psi(\lamb^{'}) =\lpotential(\lwe(\lamb^{'}))<\lpotential(\lwe(\lamb))=\Plamb$. 

Additionally, from \eqref{Psi}, we know that $\Psi(\lamb^{'})=\wpotential(\wwe(\lamb^{'})) < \wpotential(\wwe(\lamb))=\Psi(\lamb)$. Thus, the unique equilibrium edge load $\wwelamb$ necessarily changes in the direction $\zij$ in regime $\Reoneij$.

$\left[\text{Regime $\Retwoij$}\right]$: From Theorem \ref{l_behavior}, $\Lwelamb \subseteq \Ldagij$ for any $\lamb \in \Retwoij$. Since the equilibrium edge load is unique, we know $\wwelamb=\wdagij$. From \eqref{Psi} we can conclude that $\Plamb=\wpotential(\wdagij)$. Thus, $\Plamb$ as well as $\wwe(\lamb)$ remain fixed in regime $\Retwoij$.

$\left[\text{Regime $\Rethreeij$}\right]$: Following similar argument in regime $\Reoneij$, we conclude that $\Plamb$ monotonically increases in the direction $\zij$ in regime $\Rethreeij$. As a result, $\wwe(\lamb)$ changes when $\lamb$ is perturbed in the direction $\zij$ in regime $\Rethreeij$.  \QEDA

\begin{lem}{(\cite{fiacco1986convexity}, page 102)}\label{continuous_phi}
The value of the potential function in equilibrium, $\Plamb$, is convex with respect to $\lamb$ if in \eqref{eq:potential_opt}, $\potential(\q)$ is convex in $\q$, and the constraints are affine in $\q$ and $\lamb$. 
\end{lem}

\begin{lem}{(\cite{fiacco2009sensitivity}, page 3469)}\label{directional}
If in \eqref{eq:potential_opt}, the objective function $\potential(\q)$ is convex and continuously differentiable in $\q$, and additionally the set of BWE $\qwe$ and the set of Lagrange multiplies $\mu^{*}$, $\nu^{*}$ are nonempty and bounded, then $\Plamb$ is continuous and directionally differentiable in $\lamb$. Furthermore,  for any given $\i, \j \in \I$, the directional derivative of $\Plamb$ in the direction $\zij$ is $\Dplambij=\min_{\qwe \in \Qwe(\lamb)} \max_{\substack{
(\mu^{*}, \nu^{*})\\
\in \(M(\qwe), N(\qwe)\)}} \nabla_\lamb L(\qwe, \mu^{*}, \nu^*, \lamb) \zij$, 
where $M(\q^{*})$ and $N(\qwe)$ are the sets of Lagrange multipliers $\mu^{*}$ and $\nu^{*}$ in \eqref{lagrangian} associated with the BWE $\qwe \in \Qwe(\lamb)$.
\end{lem}

\emph{Proof of Proposition \ref{no_info}.}
Since the interim belief of population $\j$, $\mutj$ in \eqref{uninformed_mu} is independent with type $\tj$, the equilibrium strategy of the uninformed population $\q^{\j*}(\tj)$ must be identical across all $\tj \in \T^j$. Following \eqref{eq:mdi} and \eqref{widehatj}, the impact of information metric $\md^\j(\qwe)=\widehat{\md}^{\j}(\lwe)=0$ for any $\qwe \in \Qwe(\lamb)$, $\lwe \in \Lwelamb$ and any $\lamb$. For the sake of contradiction, we assume that the regime $\Rethreeij$ is non-empty. From Theorem \ref{l_behavior}, we know that the constraint (\ref{prime:popu_i}$_\j$) must be tight in equilibrium when $\lamb$ is in regime $\Rethreeij$. However, since $\widehat{\md}^{\j}(\lwe)=0$ for any $\lamb$, the constraint (\ref{prime:popu_i}$_\j$) is tight only when $\lamb^{j}=0$, i.e. $\lambi=1-|\lambmimj|$. This implies that the regime $\Rethreeij$ is indeed empty. Thus, there are at most two regimes $\Reoneij$ and $\Retwoij$. Following Proposition \ref{prop:relative_value}, we can conclude that $\cjlamb \geq \cilamb$.\QEDA

\begin{exam}\label{non_necessary}
We consider the game with two populations on two parallel routes ($r_1$ and $r_2$) with the following cost functions: $c_1^{\n}(\l_1)=\l_1+15$, $c_1^{\a}(\l_1)=3\l_1+15$, $c_2 (\l_2)=20\l_2+30$. The prior distribution $\theta$, the total demand $\D$, and the information environment are the same as that in Example \ref{benchmark_example}. Although both populations receive the accurate signal of the state with positive probability, we have $\lambup=1$ as the impact of information on population 2 is zero. Since the free flow travel time on $r_2$ is much higher than that on $r_1$, population 2 travelers exclusively uses $r_1$ regardless of the received signal, see Fig. \ref{fig:non_necessary}. 
\begin{figure}[H]
\centering
\begin{subfigure}[b]{0.40\textwidth}
        \includegraphics[width=\textwidth]{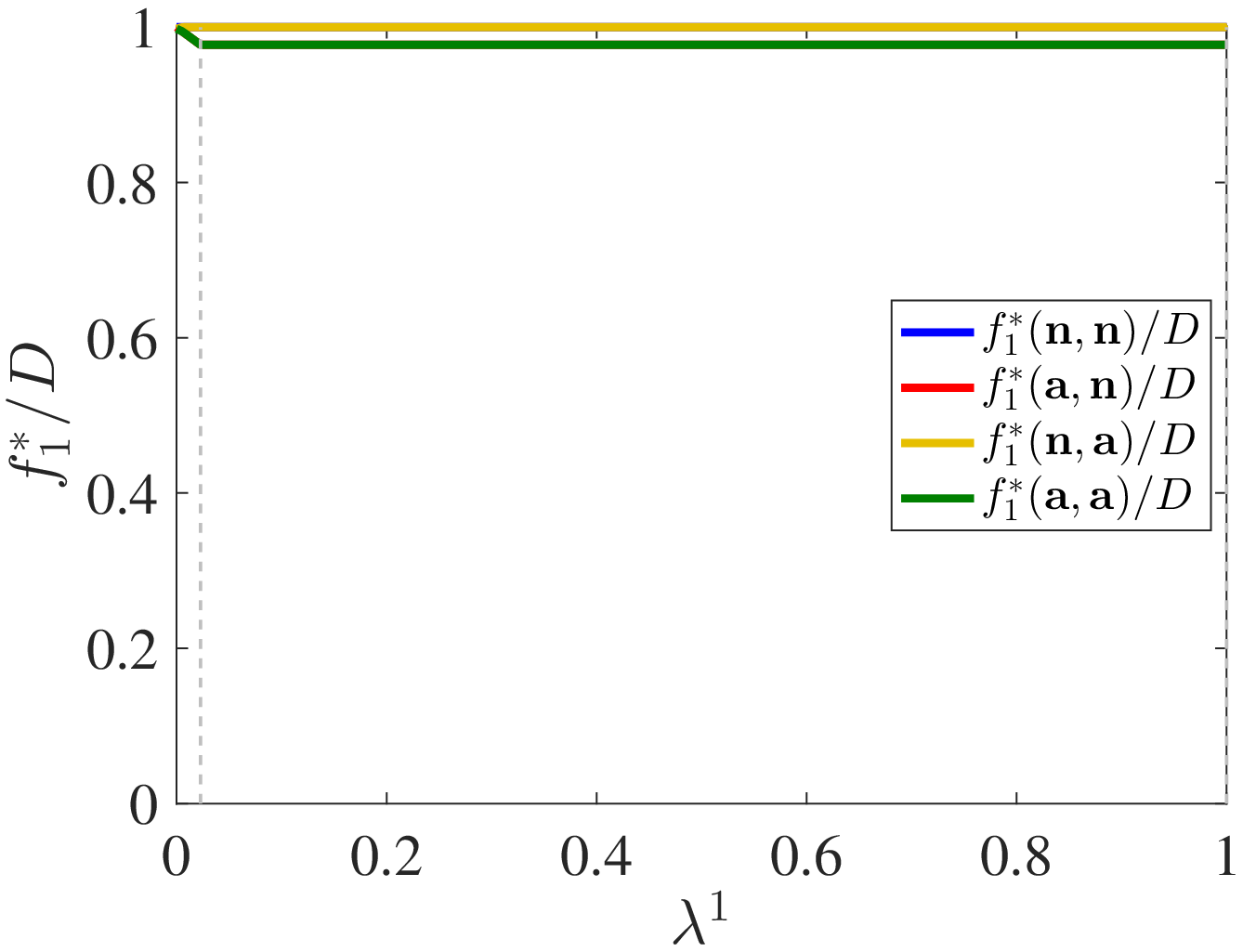}
\caption{}
    \label{fig:single_flow_unnecessary}
    \end{subfigure}
    ~ 
    \begin{subfigure}[b]{0.40\textwidth}
        \includegraphics[width=\textwidth]{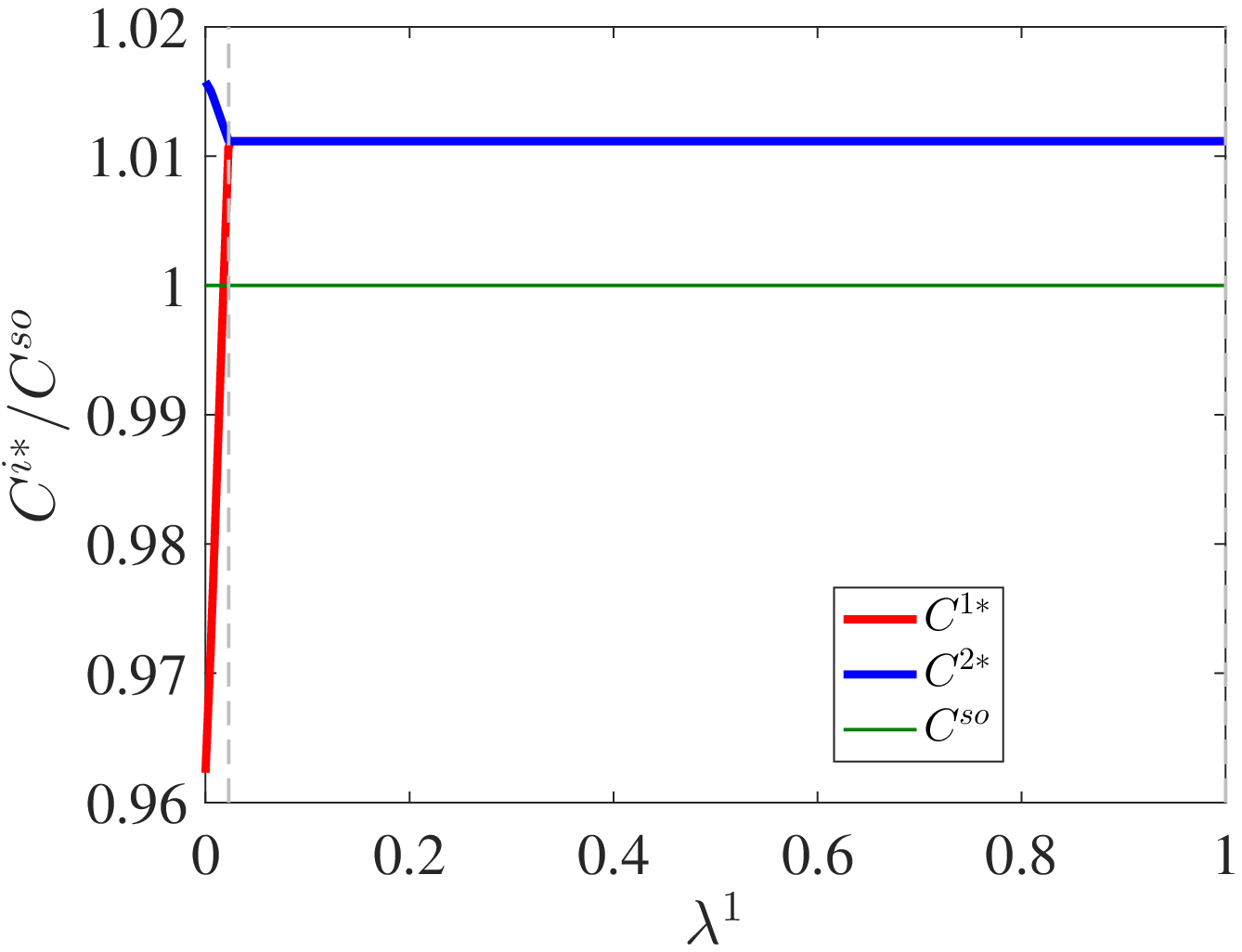}
\caption{}
    \label{fig:single_cost_unnecessary}
    \end{subfigure}
\caption{Effects of varying population sizes for Example \ref{non_necessary}: (a) Equilibrium route flows on $r_1$; (b) Equilibrium population costs.}
\label{fig:non_necessary}
\end{figure}
\end{exam}

\begin{exam}\label{single_TIS}
Consider a game with two populations on two parallel routes ($r_1$ and $r_2$). There are two states: $\{s_1, s_2\}$, each state is realized with probability 0.5. The cost functions are: $c_1^{s_1}(\l_1)=\l_1+10$, $c_1^{s_2}(\l_1)=\l_1+1$, $c_2 ^{s_1}(\l_2)=\l_2+1$, $c_2 ^{s_2}(\l_2)=\l_2+10$.  Population 1 is completely informed, and population 2 is uninformed. The total demand is $\D=1$. We now obtain that $\lambl=\lambup=1$; thus, regimes $\Retwo^{12}$ and $\Rethree^{12}$ are empty sets, and population 1 has strictly lower expected cost than population 2 for any $\lamb^1 \in (0, 1)$, see Fig. \ref{fig:strange_example}. 
\begin{figure}[H]
\centering
\begin{subfigure}[b]{0.4\textwidth}
        \includegraphics[width=\textwidth]{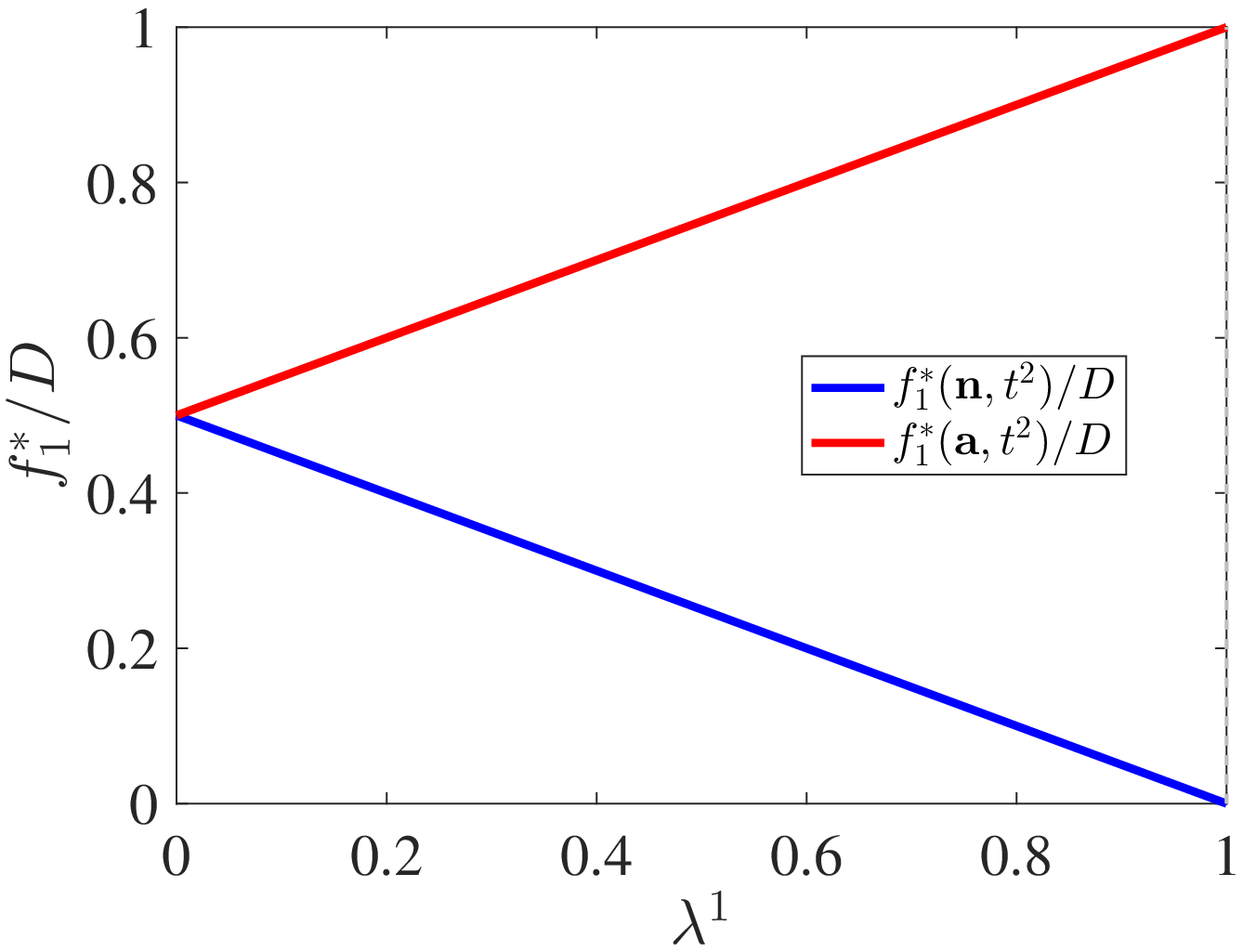}
\caption{}
    \label{fig:single_flow}
    \end{subfigure}
    ~ 
    \begin{subfigure}[b]{0.4\textwidth}
        \includegraphics[width=\textwidth]{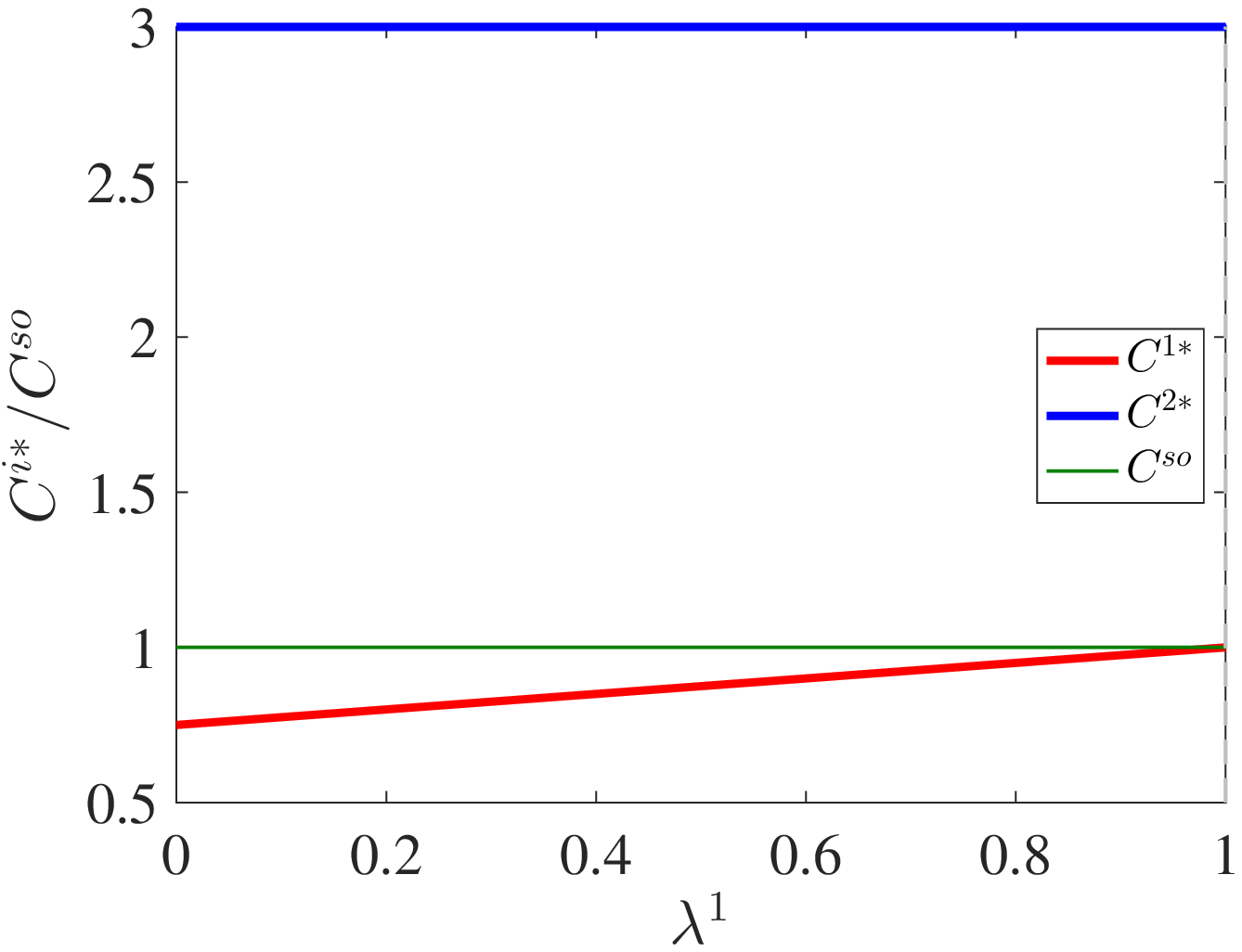}
\caption{}
    \label{fig:single_cost}
    \end{subfigure}
    \caption{Effects of varying population sizes for Example \ref{single_TIS}: (a) Equilibrium route flows on $r_1$; (b) Equilibrium population costs.}
    \label{fig:strange_example}
    \end{figure}
\end{exam}

\section{Supplementary material for Section \ref{sec:general_cost}}\label{appendix_C}
\emph{Proof of Proposition \ref{theorem:intermediate}}
Firstly, we prove that for any $\lamb \in \Lambdag$, $\Lwe(\lamb) \subseteq \Ldag$. From the definition of $\Lambdag$ in \eqref{eq:intermediate}, we know that for any $\lamb \in \Lambdag$, there exists at least one route flow $\ldag \in \Ldag$ satisfying the constraints in \eqref{opt_l}, and hence such $\ldag$ is a feasible solution of the optimization problem \eqref{opt_l}; thus $\lpotential(\ldag) \geq \Plamb$. Additionally, since $\ldag$ is an optimal solution of \eqref{eq:drop_all}, which has the same objective function as \eqref{opt_l} but without the constraints \eqref{prime:popu_i}, we conclude that $\lpotential(\ldag) \leq \Plamb$ for any feasible $\lamb$ (including $\lamb \in \Lambdag$). Thus, $\Plamb=\lpotential(\ldag)$, and $\ldag$ is an equilibrium route flow. Analogous to the argument in proof of Theorem \ref{q_opt}, the equilibrium edge load equals to $\wdag$. Since the set $\Ldag$ in \eqref{eq:Ldag_polytope} contains all route flows such that the induced edge load is $\wdag$, we can conclude that the set of equilibrium route flow $\Lwe(\lamb) \subseteq \Ldag$ for any $\lamb \in \Lambdag$. 

Next, we prove that $\Lambdag = \argmin_{\lamb} \Plamb$. We have argued in the first part of the proof that for any $\ldag \in \Ldag$, $\lpotential(\ldag) \leq \Plamb$ for any feasible $\lamb$; and since for any $\lamb \in \Lambdag$, $\Lwe(\lamb) \subseteq \Ldag$, we have $\Plamb=\lpotential(\ldag)$. Therefore, $\lpotential(\ldag)=\min_{\lamb} \Plamb$, and $\Lambdag \subseteq \argmin_{\lamb} \Plamb$. Additionally, for any $\lamb\in \argmin_{\lamb} \Plamb$, we have $\Plamb=\min_{\lamb} \Plamb=\lpotential(\ldag)$. Since $\Ldag$ includes all route flows that satisfy \eqref{sub:balance}-\eqref{sub:l_positive} and attain the minimum value of $\Plamb$, any equilibrium route flow $\lwe \in \Lwe(\lamb)$ for $\lamb \in \argmin_{\lamb} \Plamb$ must be in $\Ldag$. Hence, such $\lamb$ must also be in $\Lambdag$ defined in \eqref{eq:intermediate}, i.e. $\argmin_{\lamb} \Plamb \subseteq \Lambdag$. We can therefore conclude that $\Lambdag = \argmin_{\lamb} \Plamb$. 

From Lemma \ref{lemma:directional_derivative_psi}, we know that the function $\Plamb$ is convex in $\lamb$. Additionally, the set of feasible population size vector $\lamb$ is a closed convex set. Consequently, the set $\Lambdag = \argmin_{\lamb} \Plamb$ is convex and non-empty. 

Finally, we show that $\wwe(\lamb)=\wdag$ if and only if $\lamb \in \Lambdag$. From the first part of the proof, we know that $\Lwe(\lamb) \subseteq \Ldag$. Therefore, the unique equilibrium edge load is $\wdag$, which does not depend on $\lamb$. Additionally, for any feasible $\lamb \notin \Lambdag$, from the second part of the proof, we know that $\Plamb>\wpotential(\wdag)$. Thus, $\wwe(\lamb) \neq \wdag$.   \QEDA

\section{Extension to Networks with Multiple Origin-destination Pairs}\label{appendix_D}
In this section, we extend our model to networks with multiple origin-destination pairs, and show that all the results presented in the paper still hold. Consider a network with a set of origin-destination (o-d) pairs $\K$. Each o-d pair $k \in \K$ is connected by the set of routes $\R_k$. The set of all routes is $\R = \cup_{k \in \K} \R_k$. The demand of travelers between o-d pair $k \in \K$ is $D_k > 0$. The information environment -- state, TISs, signals and common prior --is the same as that introduced in Sec. \ref{network}. 
The fraction of travelers between o-d pair $k\in \K$ who subscribe to TIS $i \in \I$ is $\lambda_k^i$. A feasible size vector $\lambda = (\lambda_k^i)_{k \in \K, i \in \I}$ satisfies $\lambda_k^i\geq 0$ for any $k \in \K$ and any $i \in \I$, and $\sum_{i \in \I}\lambda_k^i=1$ for any $k \in \K$. We denote the strategy profile as $q=(q_{r,k}^{i}(t^i))_{r \in \R_k, i \in \I, t^i \in \Ti, k \in \K}$, where $q_{r,k}^{i}(t^i)$ is the amount of travelers in population $i$ who take route $r$ between o-d pair $k$ when the signal is $t^i$. A strategy profile $q$ is feasible if it satisfies: 
\begin{align*}
\sum_{r \in R_k} q_{r,k}^{i}(t^i)&=\lambda_k^i \cdot D_k, \quad \forall  i \in \I, \quad \forall t^i \in \T^i, \quad \forall k \in \K, \\
q_{r,k}^{i}(t^i) & \geq 0, \quad \forall r \in \R_k, \quad \forall  i \in \I, \quad \forall t^i \in \T^i, \quad \forall k \in \K. 
\end{align*}

For any feasible strategy profile $q$, the induced route flow vector is $f=(f_{r, k}(t))_{r \in \R_k, k \in \K, t \in \T}$, where $f_{r, k}(t)$ is the flow on route $r$ induced by travelers between o-d pair $k$ when the type profile is $t$: 
\begin{align}\label{eceq:q}
f_{r, k}(t)= \sum_{i \in \I}q_{r,k}^{i}(t^i), \quad \forall r \in \R_k, \quad \forall k \in \K, \quad \forall t \in \T.
\end{align}The aggregate edge load on edge $e \in \E$ when the type profile is $t$ can be written as follows: 
\begin{align}\label{we}
w_e(t)= \sum_{i \in \I}\sum_{k \in \K}\sum_{r \in \left\{\R_k|r \ni e\right\}} q_{r,k}^{i}(t^i)\stackrel{\eqref{eceq:q}}{=} \sum_{k \in \K}\sum_{r \in \left\{\R_k|r \ni e\right\}} f_{r, k}(t).
\end{align}
The expected cost of each route $r \in \R$ given type $\ti \in \Ti$, denoted $\ecrti$, can be written as in \eqref{eq:ecrti}. A feasible strategy profile $\qwe$ is a BWE if for any $k \in \K$, any $i \in \I$, and any $\ti \in \Ti$: 
\begin{align*}
\forall \r \in \R_k, \quad q_{r,k}^{i*}(t^i)>0 \quad \Rightarrow \quad \ecrtiwe \leq \ecrptiwe, \quad \forall \r' \in \R_k.
\end{align*}

We now state the extensions of our results to the network with $\K$ o-d pairs. Firstly, we can check that the following function of $\q$ is a weighted potential function of the Bayesian congestion game with $\K$ o-d pairs: 
\begin{align*}
\Phi(q)= \sum_{e \in \E}\sum_{s \in \S} \sum_{t \in \T}\pi(s, t) \int_{0}^{\sum_{i \in \I}\sum_{k \in \K}\sum_{r \in \left\{\R_k|r \ni e\right\}} q_{r,k}^{i}(t^i)} c_e^s(z) dz,
\end{align*}
Therefore, given any size vector $\lamb$, the set of equilibrium strategy profiles $\Qwe(\lamb)$ can be solved by the same convex optimization problem \eqref{eq:potential_opt} in Theorem \ref{q_opt}. The equilibrium edge load $\wwe(\lamb)$ is unique.

Secondly, with simple modifications, we characterize the set of feasible route flow vectors $\F$ as follows: 
\begin{align*}
\F \deleq \{\l ~ \vert ~\forall k \in \K, ~\text{$(\l_{r, k}(t))_{r \in \R_k, t \in \T}$ satisfies \eqref{sub:balance} -- \eqref{sub:popu_i}}\},
\end{align*}
and the set of equilibrium route flows $\Lwe(\lamb)$ can be solved by the optimization problem \eqref{opt_l} in Proposition \ref{l_opt}.
Particularly, the information impact constraint \eqref{sub:popu_i} for o-d pair $k \in \K$ and population $\i \in \I$ now becomes: 
\begin{align}\label{IIC}
D_k  - \sum_{r \in \R_k} \min_{t^i \in \T^i} f_{r,k}(t^i, t^{-i}) \leq \lambda_k^i \D_k, \quad \forall t^{-i} \in \T^{-i}, \quad \forall i \in \I, \quad \forall k \in \K. 
\end{align}
That is, the impact of information sent by TIS $i$ on the route flows between o-d pair $k$ is bounded by the amount of travelers who subscribe to TIS $i$ and travel between o-d pair $k$.

Thirdly, for any o-d pair $k$ and any pair of TISs $i$ and $j$, we can analogously analyze how the equilibrium outcomes change with the sizes $\lambda_k^i$ and $\lambda_k^j$, while the sizes of all other populations remain unchanged. We denote this direction of perturbing $\lambda$ as $z_k^{ij}$. We can show that Theorem \eqref{l_behavior} holds: three regimes (one or two may be empty) can be distinguished by whether or not the information impact all the travelers between o-d pair $k$ who subscribe to TIS $i$ (resp. $j$), i.e. whether or not \eqref{IIC} is tight at the optimum of \eqref{opt_l}.

Fourthly, the relative value of information, denoted $V_k^{ij*}(\lambda)$, is the travel cost saving experienced by travelers between o-d pair $k$ who subscribe to TIS $i$ compared with travelers who subscribe to TIS $j$. Analogous to Lemma \ref{lemma:directional_derivative_psi}, we obtain: 
\begin{align*}
V_k^{ij*}(\lambda)= C_k^{j*}(\lambda)-C_k^{i*}(\lambda)= - \frac{1}{D_k} \nabla \Phi_{z_k^{ij}}(q). 
\end{align*}
By applying sensitivity analysis, we can show that $V_k^{ij*}(\lambda)$ is positive, zero, and negative in the three regimes respectively.

Finally, we analyze how travelers between each o-d pair $k \in \K$ choose TIS subscription. By dropping the information impact constraints \eqref{IIC} in \eqref{opt_l} and following \eqref{eq:Ldag_polytope} - \eqref{eq:intermediate}, we obtain the extension of Theorem \ref{order_population}: characterization of the set $\Lambdag$, which is the set of size vectors induced by travelers' choice of TISs.  We note that how an individual traveler between o-d pair $k$ chooses information subscription not only depends on the choices of other travelers between the same o-d pair, but also the choices of travelers who travel between other o-d pairs. This is because travelers between different o-d pairs may take common routes and edges, and hence impact each other’s costs. 
\end{appendix}

\end{document}